\documentclass[superscriptaddress,english,floatfix,twocolumn,showpacs,aps,prb,preprintnumbers, reprint]{revtex4-2}

\usepackage[normalem]{ulem}
\usepackage{amsmath}
\usepackage{bigints}
\usepackage{amsfonts}
\usepackage{amssymb}
\usepackage{bm}
\usepackage{adjustbox}
\usepackage{epstopdf}
\usepackage{color}
\usepackage{textcomp}
\usepackage{gensymb}
\usepackage{graphicx}
\usepackage[dvipsnames]{xcolor}

\usepackage{dcolumn}
\usepackage{bm}
\usepackage{braket}
\usepackage{hyperref}
\usepackage{booktabs}

\usepackage{hyperref}
\hypersetup{
    colorlinks=true,
    linkcolor=BurntOrange,
    urlcolor=BrickRed,
    citecolor=Mulberry,
    pdftitle={Quantum order by disorder in Frustrated Spin Nanotubes},
    pdfauthor={Jo\~ao C. Getelina, Zekun Zhuang, Premala Chandra, Piers Coleman, Peter P. Orth, S. L. Sondhi}
}

\renewcommand{\vec}{\mathbf}
  
    \def \etal {{\it et al.}}

\definecolor{go_green}{rgb}{0.13, 0.55, 0.13}

\begin{document}
\title{Quantum Order by Disorder in Frustrated Spin Nanotubes}
\author{Jo\~ao C. Getelina}
\affiliation{Ames National Laboratory, Ames, Iowa 50011, USA}
\author{Zekun Zhuang}
\affiliation{Center for Materials Theory, Rutgers University, Piscataway, New Jersey 08854, USA}
\author{Premala Chandra}
\affiliation{Center for Materials Theory, Rutgers University, Piscataway, New Jersey 08854, USA}
\author{Piers Coleman}
\affiliation{Center for Materials Theory, Rutgers University, Piscataway, New Jersey 08854, USA}
\affiliation{Department of Physics, Royal Holloway, University of London, Egham, Surrey TW20 0EX, UK}
\author{Peter P. Orth}
\email{peter.orth@uni-saarland.de}
\affiliation{Ames National Laboratory, Ames, Iowa 50011, USA}
\affiliation{Department of Physics and Astronomy, Iowa State University, Ames, Iowa 50011, USA}
\affiliation{Department of Physics, Saarland University, 66123 Saarbr\"ucken, Germany}
\author{S. L. Sondhi}
\affiliation{Rudolf Peierls Center for Theoretical Physics, University of Oxford, Oxford OX1 3PU, UK}
\date{\today }

\begin{abstract}
We investigate quantum order by disorder in a frustrated spin nanotube formed by wrapping a $J_1$-$J_2$ Heisenberg model at 45$^\circ$ around a cylinder. Using Schwinger boson theory and Density Matrix Renormalization Group (DMRG), we have computed the ground-state phase diagram to reveal  a $\mathbb{Z}_2$ phase in which collinear spin stripes form a right or left handed helix around the nanotube. We have derived an analytic estimate for the critical $\eta_c=J_1/2J_2$ of the $\mathbb{Z}_2$-helical phase transition, which is in agreement with the DMRG results. By evaluating the entanglement spectrum and nonlocal string order parameters we discuss the topology of the $\mathbb{Z}_2$-helical phase.
\end{abstract}

\maketitle

\section{Introduction}
\label{sec:introduction}
Spin models are ``economy" strongly correlated systems where fluctuations driven by frustrated interactions result in rich physical phenomena and phase diagrams. Conceptual advances in many-body physics often arise from studies of spin models. One popular example is the emergence of order by disorder in frustrated spin models~\cite{Villain77,Villain80,Shender1982,moessnerMagnetsStrongGeometric2001}, which describes the mechanism by which quantum or thermal fluctuations lift a degeneracy of the ground state in favor of states with discrete~\cite{Chandra1990, henleySemiclassicalEigenstatesFourSublattice1998, Weber03, mulderSpiralOrderDisorder2010,zhitomirskyQuantumOrderDisorder2012,savaryOrderQuantumDisorder2012,chernDipolarOrderDisorder2013,nedicThreestatePottsNematic2022,strockozExcitonicInstabilityPottsnematic2022} or algebraic orderings~\cite{Orth12,Orth14,Jeevanesan15,Fernandes19}. 
Often studied in the context of Heisenberg spin models, order by disorder occurs in a wide range of strongly correlated systems with competing interactions~\cite{ moessnerIsingModelsQuantum2001,moessnerPhaseDiagramHexagonal2001,wesselSupersolidHardCoreBosons2005,boninsegniSupersolidPhaseHardCore2005,melkoSupersolidOrderDisorder2005b,heidarianPersistentSupersolidPhase2005b,hsiehHelicalSuperfluidFrustrated2022,greenQuantumOrderbyDisorderStrongly2018, Fernandes19}.
This physics is well understood within a renormalization group approach: the conventional scaling picture of magnetism incorporates {\sl all} fluctuations into an effective Landau-Ginzburg action for the long-wavelength modes of the spin system. Even when such an action is well-defined, high-energy, short-wavelength spin fluctuations can modify its behavior at long distances. The role of such short-wavelength fluctuations is particularly enhanced in frustrated spin systems with large ground state degeneracies. Here the associated fluctuation free energy often selects maximum entropy states that break lattice symmetries to develop discrete order.

A classic model of order by disorder is the spin-$\frac{1}{2}$ $J_1-J_2$ Heisenberg model on the two-dimensional (2D) square lattice~\cite{chandraPossibleSpinliquidState1988,ioffeEffectiveActionTwodimensional1988}. A predicted finite-temperature Ising nematic phase transition in this model~\cite{Chandra1990} has been computationally confirmed for the classical spin model several years ago~\cite{Weber03}. At the Ising nematic transition the fourfold rotation symmetry of the square lattice is broken down to a twofold rotation via the development of a local composite bond order parameter describing the relative orientation of spins. For the quantum spin-1/2 model, the transition has only recently been verified using an SU$(2)$-invariant finite temperature tensor network algorithm~\cite{Mila22}, where both quantum and thermal fluctuations are present. Indeed prior series expansion studies suggested that it might be suppressed due to quantum fluctuations~\cite{Singh03,Pierre03}. This emergent nematic transition has found analogues in other spin systems~\cite{mulderSpiralOrderDisorder2010,chernDipolarOrderDisorder2013,nedicThreestatePottsNematic2022,liCompetingEmergentPotts2022,strockozExcitonicInstabilityPottsnematic2022} 
 and unexpected realization in the iron-based superconductors~\cite{xuIsingSpinOrders2008,Fernandes-PRB-2012,Abrahams16,Kotliar22}, where it induces a nematic structural transition in the absence of long-range magnetic order. It has also been identified in several other strongly correlated materials~\cite{gopalakrishnanIntertwinedVestigialOrder2017,heckerVestigialNematicOrder2017,nieVestigialNematicitySpin2017,Fernandes19}. 

\begin{figure*}[tbh]
    \centering\hypertarget{figone}{}
    \includegraphics[width=\linewidth]{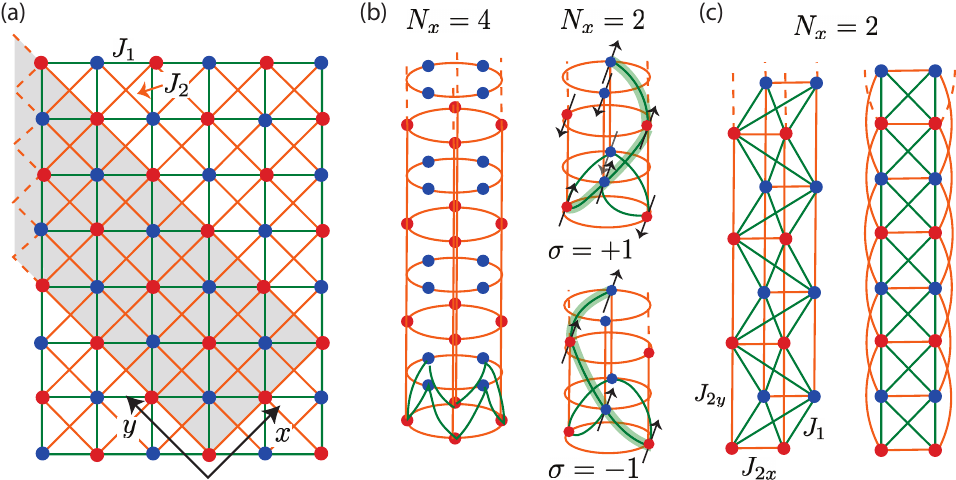}
    \caption{(a) The $J_1-J_2$ square lattice model, where $J_1$ bonds are shown in green, while $J_2$ bonds are shown in orange. In the spin nanotube, the $J_1-J_2$ model is rotated through 45$^{\circ}$ and wrapped around a cylinder as shown in (b) for $N_x=4$ and $N_x=2$ spins around the circumference. Note that this implies periodic boundary condition along the short direction of the cut and preserves the mirror symmetry between the (green) $J_1$ bonds at every site. In the $\sigma = \pm 1$ degenerate ground-states, the columns  of parallel spins along the $J_1$ bonds form left and right- handed helices around the nanotube (thick green).  (c) Showing the $N_x=2$ nanotube recast as a two-leg spin-ladder in which the $J_2$ bonds in the $x$ and $y$ direction are given values $J_{2x}$ and $J_{2y}$, respectively. } 
    \label{fig1trial}
\end{figure*}

However, the question of whether there is a zero-temperature analogue of this phenomenon in $(1+1)$D has remained open to date. Here we study such fluctuation-induced symmetry-breaking in a purely quantum spin system at zero temperature, a $J_1-J_2$ spin-$S$ Heisenberg model wrapped around a cylinder forming a frustrated spin nanotube (see Fig.~\ref{fig1trial}). Motivated by the discovery of frustrated spin chain materials~\cite{milletVanadiumIVOxide1999,gavilanoLowDimensionalSpinSystem2003,Cronin04,saha-dasguptaMathrmNa2MathrmV2005,Meissner08,Chung08,zaharkoStructuralMagneticAspects2008,Ressouche09}, spin nanotubes have been previously studied with different couplings and boundary conditions, both with and without magnetic field~\cite{luscherSolitonBindingLowlying2004,satoLegIntegerspinLadders2005,fouetFrustratedThreelegSpin2006,satoLegIntegerspinLadders2005,Sakai10,Rossini13,Oshikawa14,Pujol15,Rossini15,Shibata18}. Spin nanotubes can also be regarded as $n$-leg spin ladders (integer $n\geq 2$) with periodic boundary conditions along the short direction and we point out similarities and differences to previous studies of these systems~\cite{Dagotto92,Dagotto93, gopalanSpinLaddersSpin1994,whiteResonatingValenceBond1994,troyerThermodynamicsSpinGap1994,Scalapino00,Solyom01,Solyom08,starykhDimerizedPhaseTransitions2004}. While fluctuation induced couplings, broken symmetry phases and order by disorder have been identified in these systems (e.g. in $2$-leg ladders in  Refs.~\cite{starykhDimerizedPhaseTransitions2004, vekuaQuantumDimerPhases2006, liuExistenceDimerizedPhases2008, hikiharaPhaseDiagramFrustrated2010}), the parameter regime of the $J_1-J_2$ nanotube model we focus on here and the possibility of an emergent $\mathbb{Z}_2$-helical phase transition have not yet been investigated. 

Using complementary analytic and computational approaches, we here identify and characterize fluctuation-induced symmetry breaking in frustrated $J_1-J_2$ spin nanotubes, demonstrating that  quantum fluctuations alone can sustain transitions into phases with discrete order. Specifically, we find that quantum fluctuations select right and left handed helical states from the degenerate ground state manifold. We determine the $T=0$ quantum critical point associated with the $\mathbb{Z}_2$-helical phase to lie in the universality class of the 2D Ising model. We describe the complete ground state phase diagram for general spin length $S$ and nanotubes with different circumference $N_x$ using a Schwinger boson method~\cite{Flint2009} and for the case of spin $S=1/2$ and $N_x=2$ also using the Density Matrix Renormalization Group (DMRG)~\cite{whiteDensityMatrixFormulation1992b, schollwockDensitymatrixRenormalizationGroup2011a,Stoudenmire2012,ITensor}. We find that the results from both methods to be in good qualitative agreement. As expected, the extent of the $\mathbb{Z}_2$-helical phase is smaller in the DMRG method due to more proper treatment of quantum fluctuations but remains present. This demonstrates that gauge fluctuations beyond the large-$N$ limit used in the Schwinger boson SP($N$) approach do not destroy the $Z_2$-helical phase found in the mean-field method. 

We now discuss the modular structure of our paper. In Sec.~\ref{sec:model}, we introduce the frustrated $J_1-J_2$ nanotube model.  In Sec.~\ref{sec:SB_theory} we describe mean-field results for the phase diagrams using a symplectic large-$N$ Schwinger boson method. Complementary Density Matrix Renormalization Group (DMRG) studies for a narrow nanotube with spin $S=1/2$ are presented in Sec.~\ref{sec:dmrg_results}. We end with a discussion in Sec.~\ref{sec:discussion} where we summarize the analytic and numerical work and suggest open questions for xffuture research.

\section{Frustrated spin nanotube model}
\label{sec:model}
We consider a 1D analogue of the square lattice $J_1$-$J_2$ model shown in Fig.~\ref{fig1trial}\hyperlink{fig1trial}{(a)} and described by the Hamiltonian
\begin{equation}
    H=J_1\sum_{\langle i, j\rangle} \vec{S}_i\cdot \vec{S}_j+J_2\sum_{\langle\langle i, j\rangle\rangle} \vec{S}_i\cdot \vec{S}_j \,.
    \label{eq:Hamiltonian_J1_J2}
\end{equation}
Here, $J_1, J_2 > 0$ are fully antiferromagnetic, $\langle i,j \rangle$ sums once over first-neighbor pairs of spins (green bonds), while $\langle\langle i,j \rangle\rangle$ sums over second-neighbor pairs (orange bonds). Blue and red sites in Fig.~\ref{fig1trial} correspond to the two interpenetrating second-neighbor square sublattices coupled by $J_2$.

Due to the coexistence of $J_1$ and $J_2$ coupling, this model is frustrated, lacking a single classical configuration that simultaneously minimizes the energy of all bonds. For the square lattice model in the classical limit, it is known that the ground state consists of two decoupled antiferromagnetic sublattices when the frustration parameter
\begin{equation}
    \eta = \frac{J_1}{2J_2} < 1
\end{equation}
and exhibits N\'eel order if $\eta>1$. Here, we focus on the regime $\eta < 1$. The two antiferromagnetic sublattices are only decoupled in the absence of fluctuations, i.e. at zero temperature and for purely classical spins. Finite quantum or thermal fluctuations couple the two sublattices and select out the collinear states from the classically degenerate ground state manifold via the order by disorder mechanism~\cite{Villain77,Villain80,Shender1982,Henley1989,Chandra1990}. As noted above, the selection of one of the two collinear states breaks the fourfold rotation symmetry of the square lattice down to a twofold rotation symmetry. The transition occurs at a finite transition temperature $T_c$ via a continuous phase transition in the 2D Ising universality class~\cite{Chandra1990,Weber03,Mila22}. Heuristically, the phase transition occurs when the Heisenberg correlation length becomes comparable to the domain wall thickness separating the $\mathbb{Z}_2$ domains. 

To investigate order by disorder in $(1+1)$D, the simplest modification to the square lattice model is to wrap it around a cylinder to make a spin nanotube. To avoid breaking the  $\mathbb{Z}_2$ diagonal mirror symmetry between the two $J_1$ bond directions, it is important to rotate the strip by $45$ degrees and wrap it along the direction of the $J_2$ bonds ($x$-direction), as shown in Fig.~\ref{fig1trial}\hyperlink{fig1trial}{(a)}. On the nanotube, this symmetry corresponds to a longitudinal mirror symmetry of the lattice along the infinite direction $m_y$ that sends $y \rightarrow -y$. 
To avoid extra frustration around the tube in the $J_2 > J_1$ limit, we constrain the number of spins $N_x$ along the $x$-direction to be even. The model is expected to be gapped for even $N_x$. Since long-range order is absent at any finite temperature in 1D, we here focus on the ground state at $T=0$. The strength of quantum fluctuations is controlled by the spin length $S$ and we investigate the ground state phase diagram for different values of $S$ and frustration ratios $\eta$. As we will show below, there exists a range of frustration ratios $\eta$ where the ground state exhibits $\mathbb{Z}_2$-helical order although there is no long-range antiferromagnetic order. The emergent $\mathbb{Z}_2$-helical order breaks $m_y$ mirror symmetry and corresponds to a different chirality of the helix that follows the bonds where spins have the same relative orientation around the nanotube. This is depicted in Fig.~\ref{fig1trial}\hyperlink{fig1trial}{(b)} for the (triplet) bonds, where spins point parallel. Equivalently, one also obtains a helix if one follows the (singlet) bonds where spins point antiparallel to each other. This $\mathbb{Z}_2$-helical order is characterized by a local plaquette order parameter $\sigma$ (defined in Eq.~\eqref{eq:OP_sigma} below) that takes the two values $\sigma = \pm 1$ for the two configurations in Fig.~\ref{fig1trial}\hyperlink{fig1trial}{(b)} that are related by the longitudinal mirror symmetry $y \rightarrow -y$.

We here focus on narrow nanotubes with $N_x =2, 4$ sites along the short direction, as shown in Fig.~\ref{fig1trial}, even though our analytical theory can be straightforwardly extended to any $N_x$. For the minimal length nanotubes with $N_x = 2$, we contrast the situation of a spin nanotube, which exhibits periodic boundary conditions (PBC) along the short direction, with that of a nanostrip, which exhibits open boundary conditions (OBC). We thus consider two cases: (i) $J_{2x} = J_{2y}$, and (ii) $J_{2x} = 2 J_{2y}$. Case (i) can be regarded as a nanostrip, and case (ii) as a nanotube, where the transverse bond along the short direction is counted twice due to the wrapping around the cylinder [see Fig.~\ref{fig1trial}\hyperlink{fig1trial}{(b,c)}]. This makes the total $J_2$ coupling of a site along $x$ and $y$ bonds equal and corresponds to the situation encountered for $N_x > 2$ and the 2D square lattice, where every site has two $J_2$ neighbors along short and long directions. Importantly, we obtain qualitatively identical results for the nanostrip and nanotube geometries at $N_x = 2$. 

\section{Schwinger boson theory}
\label{sec:SB_theory}
\subsection{General formalism}
Solving spin models is a formidable task and  exact solutions are rare. A useful analytic method represents the quantum spins using Schwinger bosons, an approach which permits a natural generalization in the number of boson flavors from 2 to $N$. As $N$ becomes large, the Schwinger boson path integral is dominated by its saddle-point and the problem becomes exactly solvable in the large $N$ limit. At finite $N$ this approach provides a controlled expansion of the small parameter $1/N$ using diagrammatic methods~\cite{Auerbach2012book}. The natural extension of the Schwinger boson symmetry group of from SU(2) to SU($N$) group originally introduced by Arovas and Auerbach is limited to ferromagnets and bipartite antiferromagnets~\cite{Arovas1988}. To treat frustrated antiferromagnets, Sachdev and Read developed the SP($N$) approach~\cite{Read1991}, which however tends to underestimate the ferromagnetic correlations between the frustrated spins. In this work, we follow the symplectic-$N$ approach formalized by Flint and Coleman~\cite{Flint2009}, which treats antiferromagnetic and ferromagnetic correlations on an equal footing. 

In the symplectic-$N$ formalism, the SU(2) spin generators $S^\alpha$ are generalized to the generators $T^\alpha$ of SP($N$) ($N=2,4,...$), and the general Hamiltonian takes form
\begin{equation}
    H=\sum_{i,j}\frac{J_{ij}}{N}\vec{T}_i\cdot\vec{T}_j.
\end{equation}
The SP($N$) generators can be represented with Schwinger bosons $T^\alpha=b_\sigma^\dagger(T^\alpha)_{\sigma \sigma^\prime}b_{\sigma^\prime}$, where the spin index takes the values $\sigma=\pm 1, \pm 2, ...\pm \frac{N}{2}$. Using the completeness relation $\sum_\alpha(T^\alpha)_{ab}(T^\alpha)_{cd}=\frac{1}{4}(\delta_{ad}\delta_{bc}-\epsilon_{ac}\epsilon_{bd})$, where $\epsilon_{a b}=\text{sgn}(a)\delta_{a,-b}$, the Hamiltonian becomes
\begin{equation}
    H=\sum_{i,j}\frac{J_{ij}}{N}\left[A_{ij}^\dagger A_{ij}-B_{ij}^\dagger B_{ij}\right], \label{SchBoson}
\end{equation}
where $A_{ij}^\dagger=\frac{1}{2}\sum_\sigma b_{i\sigma}^\dagger b_{j,\sigma}$ describe normal bosonic bilinears and $B_{ij}^\dagger=\frac{1}{2}\sum_\sigma \text{sign}(\sigma) b_{i\sigma}^\dagger b_{j,-\sigma}^\dagger$ describe anomalous ones.  One can express the partition function using path integrals with the constraint $\sum_{\sigma}b_{i,\sigma}^\dagger b_{i,\sigma}=NS$ enforced by adding a Lagrange multiplier term $\exp(-\lambda_i(\sum_{\sigma}b_{i,\sigma}^\dagger b_{i,\sigma}-NS))$. After making the Hubbard-Stratonovich transformation, one obtains the saddle-point solution, which becomes exact in the large $N$ limit. This procedure is in fact equivalent to a mean-field treatment of Eq. (\ref{SchBoson}), which becomes
\begin{align}
    H_{\text{MF}}&=\sum_{i,j}\left[\Bar{h}_{ij} A_{ij}+A_{ij}^\dagger h_{ij}-\Bar{\Delta}_{ij} B_{ij}-B_{ij}^\dagger \Delta_{ij}\right.\\\nonumber
    &\left.-\frac{N}{J_{ij}}\Bar{h}_{ij}h_{ij}+\frac{N}{J_{ij}}\Bar{\Delta}_{ij}\Delta_{ij}\right]+\sum_{i,\sigma}\lambda_i(b_{i,\sigma}^\dagger b_{i,\sigma}-S)
\end{align}
with self-consistent conditions $\sum_{\sigma}\langle b_{i,\sigma}^\dagger b_{i,\sigma}\rangle=NS$, $h_{ij}=\frac{J_{ij}}{N}\langle A_{ij}\rangle$ and $\Delta_{ij}=\frac{J_{ij}}{N}\langle B_{ij}\rangle$. Alternatively, one could also determine the mean-field parameters by requiring $\partial E_g/\partial h=\partial E_g/\partial \Delta=\partial E_g/\partial \lambda=0$, where $E_g$ is the ground state energy. Since the Hamiltonian $H_\text{MF}$ contains $N/2$ equivalent copies of Kramers' doublets, it is only necessary to focus on the Hamiltonian $\tilde{H}$ of a single copy. This can be written in a compact form as
\begin{align}
\tilde{H}&=\sum_{i,j}\Psi_i^\dagger H_{ij} \Psi_j+\lambda_i(\Psi_i^\dagger \Psi_i-2S-1)\nonumber \\
&+\sum_{i,j}\left[\frac{2}{J_{ij}}\Bar{\Delta}_{ij}\Delta_{ij}-\frac{2}{J_{ij}}\Bar{h}_{ij}h_{ij}\right], \label{MFHamiltonian}
\end{align}
where we defined the Nambu spinor $\Psi_i^\dagger=\left(b_{i\uparrow}^\dagger,b_{i\downarrow}\right)$ and
\begin{equation}
    H_{ij}=\left(\begin{array}{cc}
       h_{ij} & -\Delta_{ij} \\
       \Bar{\Delta}_{ij}  & \Bar{h}_{ij} 
    \end{array}\right).
\end{equation}

\subsection{Mean-field phase diagrams}
\begin{figure}[tbh]
    \centering
    \includegraphics{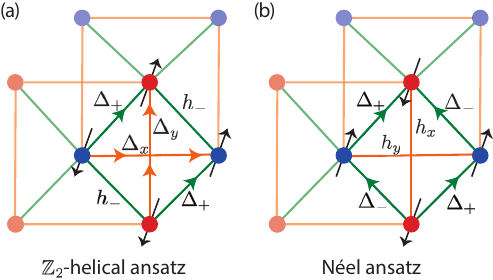}
    \caption{Schwinger boson mean-field ans\" atze used in this work, where $\Delta$ defines the strength of the  (antiferromagnetic) pairing and $h$ the strength of the (ferromagnetic) particle-hole hybridization. Since  $\Delta_{ij} = - \Delta_{ji}$ along the antiferromagnetic bonds, to avoid ambiguity the arrow points to the designated ``$i$" site on the bond.  (a) $\mathbb{Z}_2$-helical ansatz and (b) N\'eel ansatz.  }
    \label{fig:SB_ansaetze}
\end{figure}

In principle, the mean-field states of Eq. (\ref{eq:Hamiltonian_J1_J2}) can be obtained by minimizing the ground state energy of Eq.~\eqref{MFHamiltonian} with respect to all possible mean-field parameters, which is generally rather difficult. Here we only consider translationally invariant mean-field ans\"atze that are connected via second-order phase transitions to either the $\mathbb{Z}_2$ ansatz or the N\'eel ansatz, which are shown in Fig.~\ref{fig:SB_ansaetze}. The first-order phase transition is then determined by directly comparing the ground state energy of different states, while the continuous phase transition is determined by analyzing the instability of ground state energy through the calculation of its second derivatives. Below we show our final results and the detailed calculations can be found in Appendix~\ref{appendix:SchBoson}.

\begin{figure}[tbh]
    \centering
    \includegraphics[width=\linewidth]{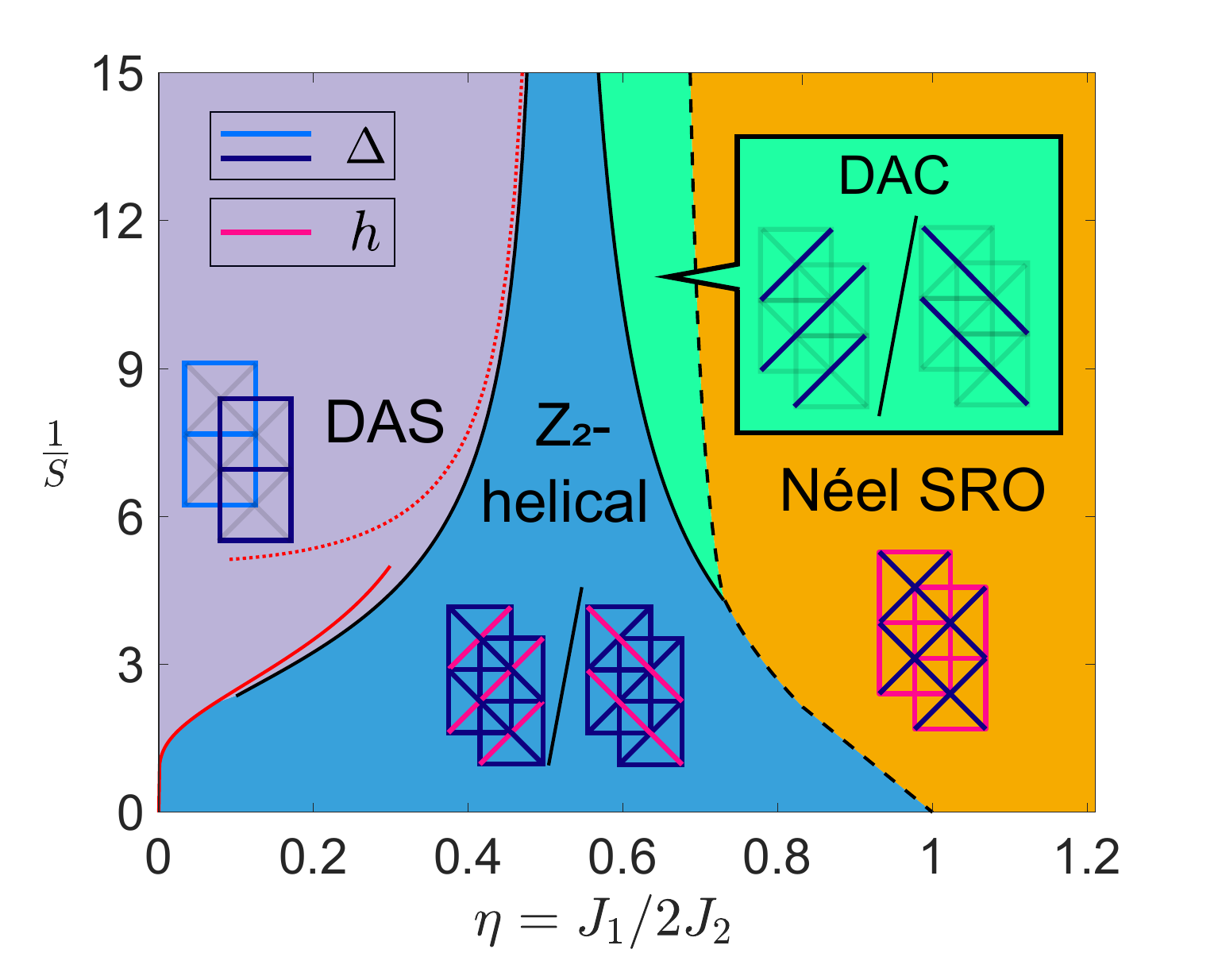}
    \caption{The phase diagram of $N_x=4$ spin nanotube predicted by Schwinger boson theory. The black solid (dashed) line represents the second(first)-order phase transition, while the the red solid (dotted) line represents the $\mathbb{Z}_2$ critical line in the one(two)-dimension limit. Also shown are the valence bond structures of the corresponding mean-field ans\"atze, where blue lines represent antiferromagnetic bonds while the violet-red lines denote ferromagnetic bonds (same for Fig. \ref{fig:Nx2phasediagram_2} and Fig. \ref{fig:Nx2phasediagram_1}). Here, DAS and DAC refer to decoupled antiferromagnetic sublattices and chains, respectively. }
    \label{fig:Nx4diagram}
\end{figure}

Figure~\ref{fig:Nx4diagram} shows the mean-field phase diagram of the $N_x=4$ spin nanotube. For $0<\eta<1$, the classical ground state of the system contains two decoupled antiferromagnetic sublattices (DAS). When $S$ is large but finite, the quantum fluctuations remove the classical degeneracy and lead to $\mathbb{Z}_2$-helical long-range order (LRO), which corresponds to a different handedness of the spin texture around the nanotube (see Fig.~\ref{fig1trial}{(b)}). When either $\eta$ or $S$ decrease, quantum fluctuations are enhanced and the order melts such that the system continuously transitions to the DAS phase. At intermediate $\eta$, there exists a ``decoupled antiferromagnetic chain" (DAC) phase, which exists at unrealistically small $S < 1/2$ and may be unstable when finite $N$ fluctuations are considered. We note that this regime can be reached when considering SU($N$) spins with $N>2$ for which quantum fluctuations are enhanced and small spin $S\sim 1/N$ is possible~\cite{Arovas1988,Read1991,Flint2009}. For sufficiently large $\eta$, antiferromagnetic alignment of nearest-neighbor spins is favored, and the system enters N\'eel short-range order (SRO) phase through a first-order transition. 

For general $N_x\geq 4$, the phase diagram is expected to be similar to the $N_x=4$ case. Near the $\mathbb{Z}_2$ phase transition, the system may be regarded as one-dimensional if the spin correlation length is much larger than the circumference of the nanotube ($\eta\rightarrow 0$), or two-dimensional if in the opposite limit ($\eta\rightarrow 1/2$). The asymptotic solution of the second-order phase transition in the 1D regime is given by
\begin{equation}
    S_\text{$\mathbb{Z}_2$}\exp\left(-\frac{N_x\pi(S_{\text{$\mathbb{Z}_2$}}+1/2)}{\sqrt{2}}\right)=\frac{\sqrt{\gamma_N}\eta}{16N_x\pi }, \label{Nx4SchMain}
\end{equation}
where $\gamma_N$ is a $N_x$-dependent constant (see Appendix \ref{appendix:SchBoson}). In the 2D regime, the critical line for the spin nanotube becomes asymptotic to that in the exact two dimensions. From Fig.~\ref{fig:Nx4diagram} one can see that when $\eta$ gets closer to zero, the analytic solution in Eq.~\eqref{Nx4SchMain} agrees better with the exact numerical result; while when $\eta$ is close to $1/2$, the difference of $S_{\text{Z2}}$ between the two-dimensional system ($N_x=\infty$) and one-dimensional nanotube (finite $N_x$) vanishes. One expects that when $N_x$ increases, a larger range of $\mathbb{Z}_2$ transition line will coincide with that of the two-dimensional system, while a narrower range of $\mathbb{Z}_2$ transition can be described by the Eq. (\ref{Nx4SchMain}), and the system makes crossover from 1D to 2D at a smaller value of $\eta$.

\begin{figure}[t!]
    \centering
    \includegraphics[width=\linewidth]{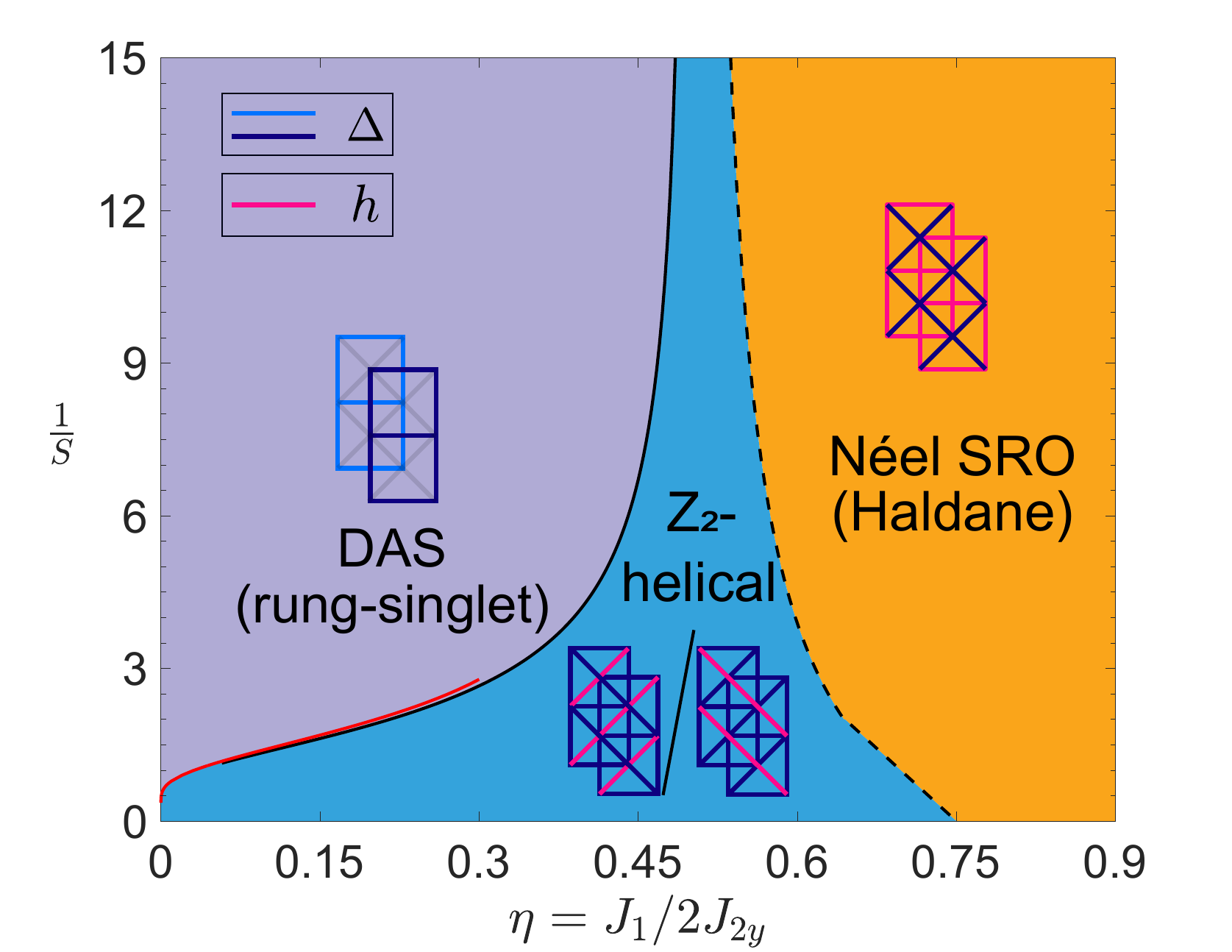}
    \caption{The phase diagram of $N_x=2$ spin nanostrip, i.e. case (i) $J_{2x} = J_{2y}$, predicted by Schwinger boson theory. The black solid (dashed) line represents the second(first)-order phase transition, while the red solid line is the analytic solution Eq. (\ref{Sch2}).}
    \label{fig:Nx2phasediagram_2}
\end{figure}

We now demonstrate the special case $N_x=2$ where the system may be regarded as a two-leg ladder. For the spin nanostrip ($J_{2x} = J_{2y}$), the phase diagram shares many similarities with that of the $N_x=4$ case, as shown in Fig. \ref{fig:Nx2phasediagram_2}. The critical line when $\eta\rightarrow 0$ is given by the asymptotic solution
\begin{equation}
    S_{\text{$\mathbb{Z}_2$}}\exp\left(-\frac{2\sqrt{6}\pi(S_{\text{$\mathbb{Z}_2$}}+1/2)}{3}\right)=\frac{\sqrt{\gamma_2^\prime}\eta}{32\pi}, \label{Sch2}
\end{equation} 
where $\gamma_2^\prime=8\left[\frac{2}{3}-\sqrt{\frac{2}{3}}\ln{(\sqrt{2}+\sqrt{3})}\right]\approx 2.15$. For $S=1/2$ the $\mathbb{Z}_2$ order phase exists when $0.22<\eta<0.65$, while for $S=1$ the $\mathbb{Z}_2$ order phase exists when $0.03<\eta<0.7$.

\begin{figure}[t!]
    \centering
    \includegraphics[width=\linewidth]{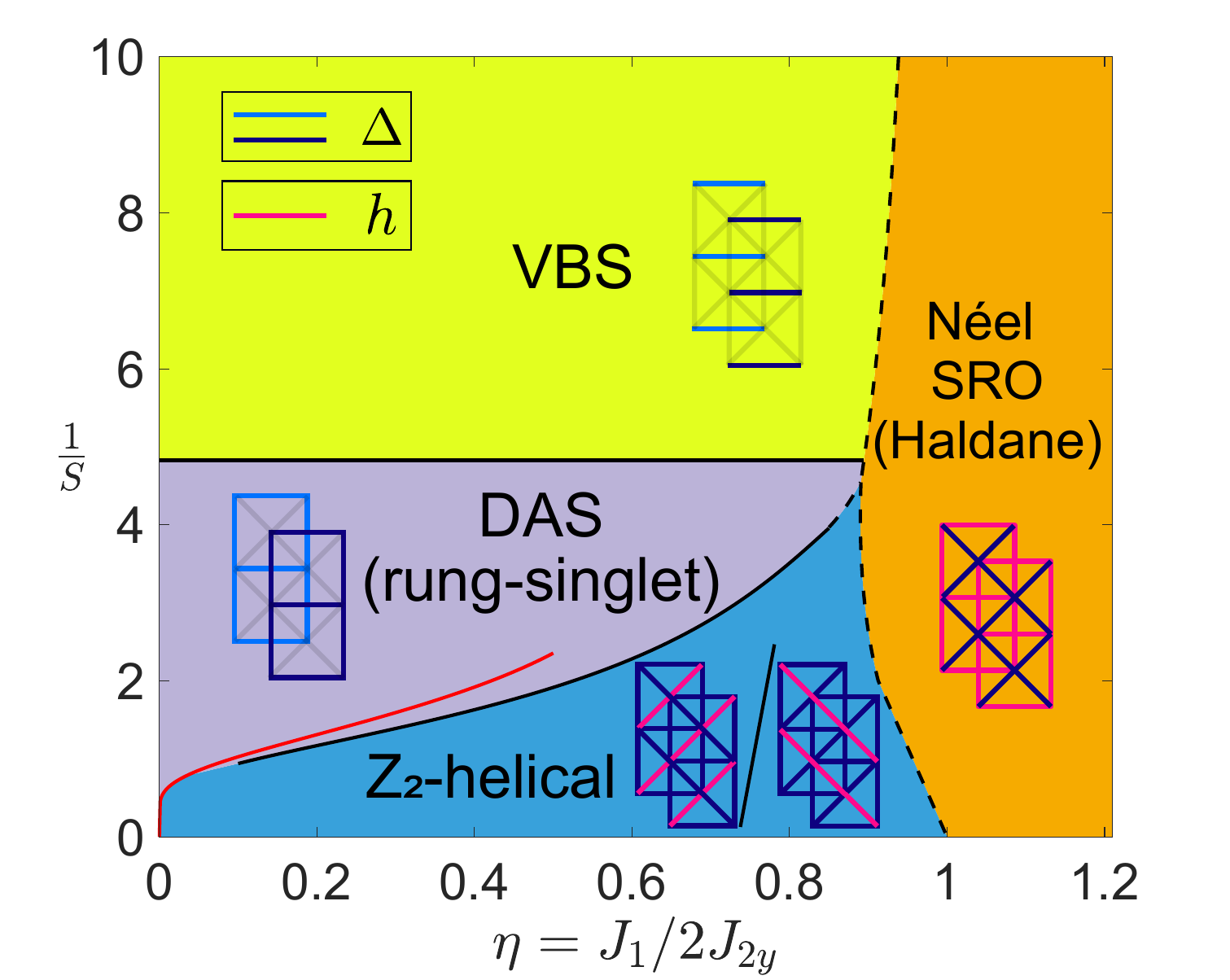}
    \caption{The phase diagram of $N_x=2$ spin nanotube, i.e. case (ii) $J_{2x} = 2 J_{2y}$, predicted by Schwinger boson theory. The black solid (dashed) line represents the second(first)-order phase transition, while the red solid line is the analytic solution Eq.~\eqref{Sch}. Here VBS refers to a valence bond solid phase.}
    \label{fig:Nx2phasediagram_1}
\end{figure}

The phase diagram of the $N_x=2$ spin nanotube ($J_{2x} = 2J_{2y}$) is somewhat distinct from that of the other cases, as seen from Fig. \ref{fig:Nx2phasediagram_1}. The main difference is that there also exists a valence bond solid (VBS) phase, which occurs only for unrealistically small spin length $S<0.21$. The emergence of this phase can be understood by noting that, the $J_2$ coupling between two sites in the $x$ direction is twice larger than those in the $y$ direction, which hence favors the formation of transverse valence bonds. Nevertheless, the asymptotic solution of $\mathbb{Z}_2$ critical line has a similar form
\begin{equation}
    S_{\text{$\mathbb{Z}_2$}}e^{-\sqrt{2}\pi (S_{\text{$\mathbb{Z}_2$}}+1/2)}=\frac{\sqrt{\gamma_2}\eta}{32\pi}, \label{Sch}
\end{equation}
where $\gamma_2=8(\sqrt{2}\ln{(\sqrt{2}+1)}-1)\approx 1.97$. The $\mathbb{Z}_2$ order phase exists when $0.53<\eta<0.91$ for spin-$1/2$ system, while it exists when $0.12<\eta<0.96$ for spin-1 system.

Our results in Eqs.~\eqref{Nx4SchMain}-\eqref{Sch} may be understood in a similar way as done by Chandra \etal{} for the $\mathbb{Z}_2$ transition driven by thermal fluctuation in two dimensions~\cite{Chandra1990}: the $\mathbb{Z}_2$ order exists when the 1D Heisenberg correlation length $\xi$ is larger than the $\mathbb{Z}_2$ domain wall width $w_{\text{DW}}$, and a simple estimate gives $\xi\sim \exp(N_x\pi(S+1/2)/\sqrt{2})$, $w_{\text{DW}}\sim J_2/J_1$. Therefore, the $\mathbb{Z}_2$ phase transition occurs when $\exp(-N_x\pi(S+1/2)/\sqrt{2})\sim \eta$. Note that compared to this estimate, our results (see Eq.~\eqref{Nx4SchMain}) based on Schwinger boson approach has an extra spin-dependent factor, which has also been seen in a similar Schwinger boson calculation of 2D $\mathbb{Z}_2$ phase transition at finite temperature \cite{Flint2009}. We speculate such difference is due to the negligence of finite $N$ fluctuation in the mean-field approach and expect such difference to be eliminated when the fluctuations are taken into account.

\section{DMRG results}
\label{sec:dmrg_results}
\subsection{Model and simulation parameters}
\label{subsec:dmrg_model_and_simulation_parameters}
In this section we present DMRG simulation results for the spin-$1/2$ nanotube. We focus on nanotubes with the minimal length $N_x =2$ along the short direction, as shown in Fig.~\ref{fig1trial}\hyperlink{fig1trial}{(b,~c)}. In a different parameter regime, this model has been investigated in Refs.~\cite{vekuaQuantumDimerPhases2006, liuExistenceDimerizedPhases2008}, where symmetry-breaking columnar dimer (CD) and staggered dimer (SD) phases were reported to exist between the rung-singlet and the Haldane phase. We here show that a $\mathbb{Z}_2$-helical phase emerges when the coupling across the nearest-neighbor bonds is equal to the diagonal bond coupling. This parameter choice arises naturally in our construction of wrapping the $J_1$-$J_2$ square lattice model around a cylinder, since both are $J_1$-bonds then. 

\begin{figure}[t!]
    \centering  \includegraphics{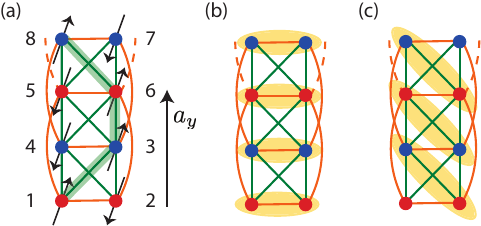}
    \caption{(a) Site labeling used in DMRG calculations. Order parameter $\sigma_i$  of the $\mathbb{Z}_2$-helical phase explicitly reads for the first double plaquette $\sigma_{1} = (\mathbf{S}_3 - \mathbf{S}_4) \cdot (\mathbf{S}_1 - \mathbf{S}_2 + \mathbf{S}_6 - \mathbf{S}_5)$. Thick green bonds follow parallel spins. We use the unit cell vector $a_y$ such that there are four spins per unit cell. The total number of spins in the model is given by $N = 4 L_y$, where $L_y$ is the total number of plaquettes. (b) Pairing of spins-1/2 in the odd string order parameter $\mathcal{O}^z_\text{odd}$ in Eq.~\eqref{eq:odd_string_op} (c) Pairing of spins in the even string order parameter $\mathcal{O}^z_\text{even}$ in Eq.~\eqref{eq:even_string_op}.}
    \label{fig:ops}
\end{figure}

We choose a unit cell vector $a_y$ such that the 1D model contains four basis sites per unit cell, two ``blue" and two ``red" sites [see Fig.~\ref{fig:ops}(a)]. As mentioned before, blue and red sites correspond to the two interpenetrating second-neighbor square sublattices coupled by $J_2$. We denote the total number of unit cells by $L_y$, and the model thus contains $N = 4 L_y$ spins. We consider two values for the ratio $J_{2x}/J_{2y}$, where $J_{2x}$ describes the coupling along the rungs, and $J_{2y}$ is the second-neighbor coupling along the chains. We refer to case (i) $J_{2x} = J_{2y}$ as the nanostrip, and to case (ii) $J_{2x} = 2 J_{2y}$, where the transverse bond along the short direction is counted twice due to the wrapping around the cylinder. 

We use the Julia simulation library iTensor~\cite{ITensor,ITensor-r0.3} to perform the DMRG calculations in this work. For each DMRG run we allow for up to $30$ sweeps, with only the final sweep being truncated at a discarded weight of maximally $10^{-10}$. The maximum bond dimension $m$ is doubled after every three sweeps, and we set a final value of $m = 3840$ for the final truncated sweep. We consider open BCs along the long direction with a total number of $L_y$ unit cells. The labeling of the sites follows a snake-like pattern, as shown in Fig.~\ref{fig:ops}\hyperlink{fig:ops}{(a)}. 

The DMRG simulations are set to be quantum-number conserving, which reduces the simulation time. For spin chains the conserved quantum number corresponds to the $z$-component of the total spin $S^z_{\textrm{tot}}$. 
When computing the spin gap, we perform two DMRG runs for each system size $L_y$, one in the sector with $S^z_{\textrm{tot}}=0$, and  another one in the sector with $\left|S^z_{\textrm{tot}}\right|=1$. After determining the ground state in each total spin sector, we also find the corresponding first-excited state from a second DMRG run, where we add a penalty term to the Hamiltonian that is the weighted projection to the previously determined ground state. 

To directly observe the emergence of $\mathbb{Z}_2$-helical order we measure a suitable plaquette bond order parameter $\sigma$. 
For the site labeling in Fig.~\ref{fig:ops}\hyperlink{fig:ops}{(a)}, we define the order parameter as 
\begin{equation}
    \sigma \left( L_y \right) = \frac{1}{8 (L_y - 1) S (S+1)} \sum_{i=1}^{L_y - 1}\sigma_i,
    \label{eq:OP_sigma}
\end{equation}
where the bond order parameter at plaquette $i$ reads
\begin{equation}
    \sigma_i = \left( \vec{S}_{4i-1} - \vec{S}_{4i} \right) \cdot \left( \vec{S}_{4i-3} + \vec{S}_{4i+2} - \vec{S}_{4i-2} - \vec{S}_{4i+1} \right)\,.
    \label{eq:OP_sigma_i}
\end{equation}
Here, the spin operators can be expressed in terms of Pauli matrices as $\vec{S}_i = \frac12 (X_i, Y_i, Z_i)$.
We choose a convenient normalization in Eq.~\eqref{eq:OP_sigma} such that $\sigma = 1/3$ in the $Z$ basis N\'eel product state $\ket{\uparrow \downarrow \uparrow \downarrow \downarrow \uparrow \downarrow \uparrow \cdots}$, where the sites are increasing from left to right starting from $i=1$ and $Z\ket{\uparrow} = \ket{\uparrow}$, $Z\ket{\downarrow} = -\ket{\downarrow}$.
Also note that the summation in Eq.~\eqref{eq:OP_sigma} only runs until $i=L_y-1$, because we use a lattice geometry with an equal number of ``blue" and ``red" sites, and the top row of spins is thus ``blue". It can be shown straightforwardly that this order parameter $\sigma \sim (\Delta_+^2-\Delta_-^2+h_-^2-h_+^2$) in the Schwinger boson theory (see Sec. \ref{sec:SB_theory}), thus a nonzero $\sigma$ indicates the $\mathbb{Z}_2$-helical phase identified in the previous section.

The initial state of the DMRG simulations is chosen to be a $Z$ basis N\'eel product state for each sublattice. If the first site of the ``blue" sublattice (site $i = 3$) has the same spin direction as the first site of the ``red" sublattice (site $i = 1$), i.e., $\ket{\uparrow \downarrow \uparrow \downarrow \downarrow \uparrow \downarrow \uparrow \cdots}$, the initial state has $\sigma=1/3$. In contrast, if they are in antiparallel directions, i.e., $\ket{\uparrow \downarrow \downarrow \uparrow \downarrow \uparrow \uparrow \downarrow \cdots }$, then we find $\sigma=-1/3$. In cases where $\sigma \neq 0$, we have verified that the final converged state at the end of the DMRG simulation shares the same $\sigma$ sign as the initial state. By preparing different N\'eel product initial states, we have also explicitly checked that the two states with opposite $\sigma$ signs are energetically degenerate. However, the DMRG method can still return a state that is a superposition of the two possible symmetry breaking states. In order to avoid converging into a superposition of symmetry-broken states, we added an external symmetry-breaking pinning field at the top and bottom boundary plaquettes
\begin{align}
    H_\lambda = \lambda & \left[ \sigma_1 + \left( \vec{S}_{N-3} - \vec{S}_{N-2} \right) \right. \\\nonumber
    & \left. \cdot \left( \vec{S}_{N-5} + \vec{S}_{N} - \vec{S}_{N-4} - \vec{S}_{N-1} \right) \right]\,.
\end{align}
In the following, we use a pinning field of strength $\lambda$ = 1, except where explicitly noted. To still be able to observe the behavior free of the pinning field, one can apply the subtraction method as described in Ref.~\cite{Stoudenmire2012}.

\begin{figure*}[tbh]
    \centering
    \includegraphics[width=\linewidth]{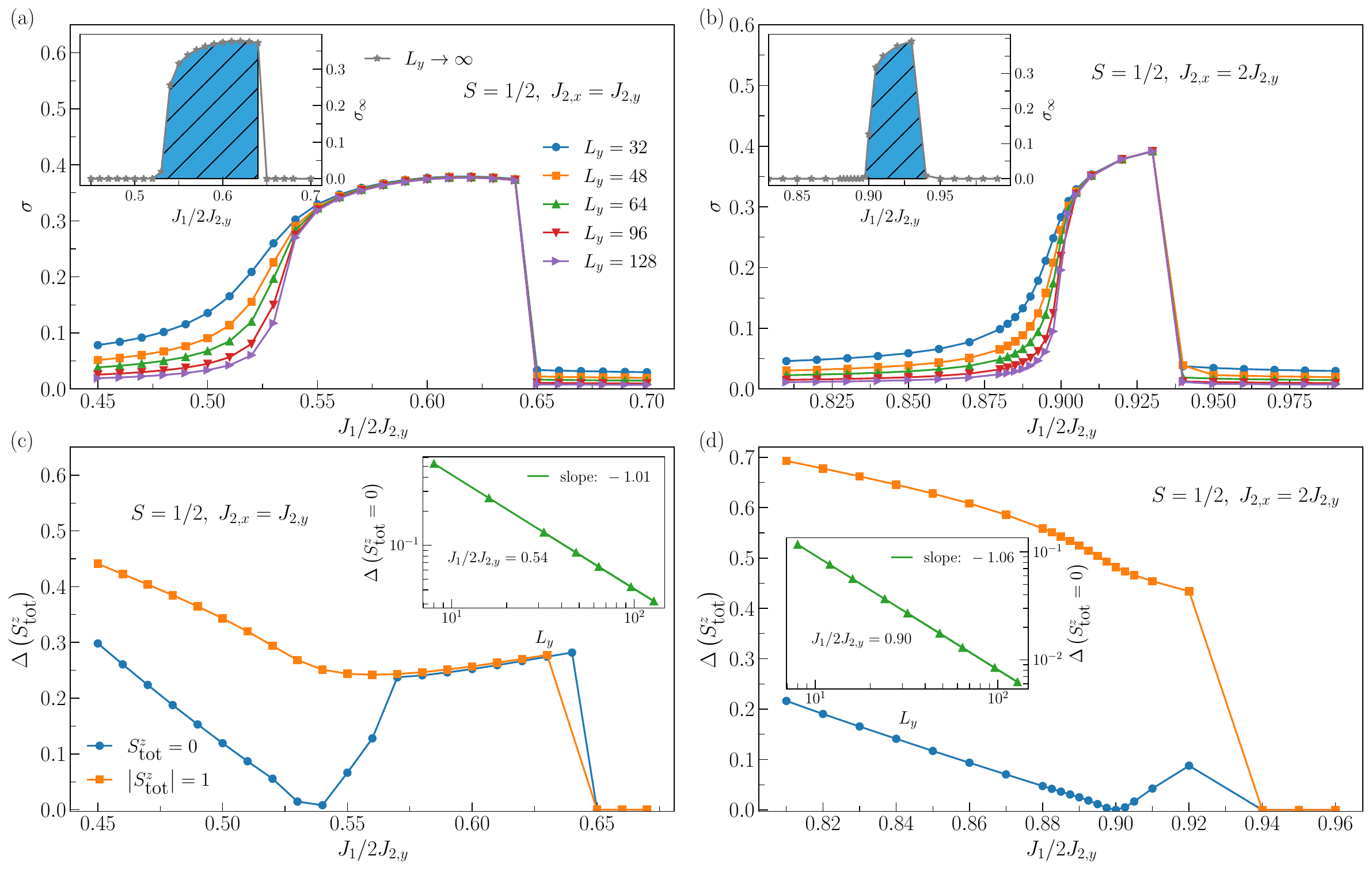}
    \caption{(a, b) $\mathbb{Z}_2$-helical order parameter $\sigma$ as a function of $\eta = J_1/2 J_2$ for $S = 1/2$, $N_x = 2$ and various cylinders of length $L_y$. Here, $L_y$ refers to the total number of unit cells such that the total number of spins equals $N = 4L_y$. The inset shows the value of  $\sigma$ extrapolated to $L_y \rightarrow \infty$. (c, d) Bulk system energy difference $\Delta$ between the ground and first-excited states in the sector $S_\textrm{tot}^z = 0$ (circles), and between the ground states of sectors $\left|S_\textrm{tot}^z \right|=1$ and $S_\textrm{tot}^z= 0$ (squares). Left and right-hand side panels correspond to cases $J_{2x} = J_{2y}$ and $J_{2x} = 2J_{2y}$, respectively.}
    \label{fig:dmrg-op-and-gap}
\end{figure*}
\subsection{Phase diagram and energy gap}
In this section, we determine the $\mathbb{Z}_2$-helical order parameter and the energy gap above the (possibly degenerate) ground state as a function of $J_1/(2J_{2y})$. This directly establishes the presence of a $\mathbb{Z}_2$-helical phase with a nonzero local order parameter in the system. 
In Fig.~\ref{fig:dmrg-op-and-gap}\hyperlink{fig:dmrg-op-and-gap}{(a,~b)}, we show the order parameter $\sigma$ in the converged DMRG ground state for different $\eta = J_1/(2J_{2y})$ and system sizes up to $L_y = 128$. Panel (a) corresponds to the nanostrip case $J_{2x} = J_{2y}$, and (b) to the nanotube $J_{2x} = 2 J_{2y}$. We obtain our results starting from an initial N\'eel product state $\ket{\uparrow \downarrow \cdots}$ with $\sigma = 1/3$ and in the presence of a boundary pinning field $\lambda = 1$. The inset displays the order parameter in the thermodynamic limit, which is obtained by extrapolating the results $\sigma(L_y)$ to $1/L_y \rightarrow 0$. The blue shaded area thus represents the $\mathbb{Z}_2$-helical phase, which is characterized by a nonzero local order parameter $\sigma_\infty$. We note that we are using the ``subtraction method" formula~\cite{Stoudenmire2012}, which reads
\begin{align}
    \sigma_\infty ( L_y^{(1)}, L_y^{(2)}) = \frac{1}{L_y^{(2)} - L_y^{(1)}} & \left[ (L_y^{(2)}-1)\sigma(L_y^{(2)}) \right. \\\nonumber
    & \left. - (L_y^{(1)}-1)\sigma(L_y^{(1)}) \right]\,,
\end{align}
to remove boundary effects in the finite $L_y$. Comparing panels (a) and (b), we observe that the $\mathbb{Z}_2$-helical phase is wider and centered at a smaller value of $\eta$ for the nanostrip case $J_{2x} = J_{2y}$ (panel a). This is consistent with our results using Schwinger boson theory (cf. Figs.~\ref{fig:Nx2phasediagram_2} and~\ref{fig:Nx2phasediagram_1}).

The $\mathbb{Z}_2$-helical phase extends over a range $\eta_{c,1} \leq \eta \leq \eta_{c,2}$. We find the phase boundary on the left to be at $\eta_{1,c} = 0.540(5)$ for the nanostrip and $\eta_{c,1} = 0.902(5)$ for the nanotube. The boundary on the right occurs at $\eta_{c,2} = 0.650(5)$ for the nanostrip and at $\eta_{c,2} = 0.940(5)$ for the nanotube.
While the order parameter smoothly vanishes at $\eta_{c,1}$, it abruptly drops to zero at $\eta_{c,2}$. This suggests the presence of a continuous phase transition at $\eta_{c,1}$ and a first order transition at $\eta_{c,2}$. This will be confirmed by our detailed numerical study below, where we also show that the continuous phase transition at $\eta_{c,1}$ lies in the universality class of the 2D Ising model (see Sec.~\ref{subsec:Ising_criticality}). 

In Figs.~\ref{fig:dmrg-op-and-gap}\hyperlink{fig:dmrg-op-and-gap}{(c, d)}, we present the energy gap above the ground state as a function of $\eta$. Panel (c) is for the nanostrip and panel (d) is for the nanotube. Before discussing the results, it is important to note that the degeneracy of the ground state changes as we vary $\eta$. While the ground state is singly degenerate for $\eta < \eta_{c,1}$, we find it to be twofold degenerate within the $\mathbb{Z}_2$-helical phase for $\eta_{c,1} < \eta < \eta_{c,2}$. In both cases the ground state lies in the sector $S^z_\text{tot} = 0$. Finally, it is fourfold degenerate for $\eta > \eta_{c,2}$ with two ground states lying in the $S^z_{\text{tot}} = 0$ sector and the other two in the $|S^z_{\text{tot}}| = 1$ sectors. 

Figs.~\ref{fig:dmrg-op-and-gap}\hyperlink{fig:dmrg-op-and-gap}{(c, d)} contain two energy gap curves: one is the energy gap between the first-excited and the ground state of sector $S^z_\textrm{tot}=0$ (blue), and the other one is the usual spin gap, which gives the energy difference between the ground states of sectors $\left|S^z_\textrm{tot}\right|=1$ and $S^z_\textrm{tot}=0$ (orange). 
The shown gap values are obtained by fitting the data for various cylinder lengths $L_y$ to the equation $f\left(\mathbf{a}, L_y\right) = \sum_{i=0}^3 a_i \left(L_y\right)^{-i}$. We show a scaling plot of the spin gap $\Delta(|S^z_{\text{tot}}| = 1)$ for several $\eta$ values in Appendix~\ref{appendix:scaling}.

We find a nonzero energy gap above the doubly degenerate ground state in the $\mathbb{Z}_2$-helical phase. The gap to the first excited state in the $S^z_{\text{tot}} = 0$ sector closes smoothly at $\eta_{c,1}$. The insets show the (singlet) gap at $\eta_{c,1}$ versus system size, which follows a power law $\Delta(S^z_{\text{tot}} = 0) \propto L_y^{-z}$ with exponent $z = 1$. This is consistent with the 2D Ising universality class. The gap reopens for $\eta < \eta_{c,1}$, where we find a singly degenerate ground state. The gap to the triplet sector stays nonzero across the critical point. We note that due to SU(2) symmetry, all triplet energies are threefold degenerate, which explains the degeneracy of the blue and orange curves inside the $\mathbb{Z}_2$-helical phase in Fig.~\ref{fig:dmrg-op-and-gap}\hyperlink{at $\eta_{c,1}$}{(c)}. 

In contrast, at the first order transition $\eta_{c,2}$ we find that both the singlet and the triplet gap abruptly vanish and the ground state becomes fourfold degenerate. We have checked that it remains gapped above the ground state by calculating the second excited state in the $S^z_{\text{tot}} = 0$ and $|S^z_{\text{tot}}| = 1$ sectors (not shown). 

To better understand the two phases that surround the $\mathbb{Z}_2$-helical phase, let us investigate the limits $\eta \ll 1$ and $\eta \gg 1$. For $\eta = 0$ (or equivalently $J_1 = 0$), the $J_1-J_2$ spin-$1/2$ nanotube model decomposes into two nearest-neighbor Heisenberg two-leg ladders of length $L_y$. This model has been studied extensively in the literature~\cite{Dagotto92,Dagotto93, whiteResonatingValenceBond1994,Tsvelik96,Pujol97}, and its ground state is well-known to be a resonating-valence-bond (RVB) state. The RVB state consists of a a superposition of singlet coverings with nearest-neighbor singlets having larger weights. Nearest-neighbor singlets occur either along the leg or across the rungs of the ladder, and this phase is usually called the ``rung singlet" (RS) phase. 

As we start from the RS state at $\eta=0$ and gradually increase $J_1$, the ``red" and ``blue" sublattices are reconnected, and we expect that singlets on $J_1$ bonds now become increasingly important, i.e., diagonal singlets and nearest-neighbor singlets along the chains. Since the gap remains open and the ground state remains singly degenerate, we remain in the RS phase for $\eta < \eta_{c,1}$. A similar behavior was reported, for example, in Ref.~\cite{Solyom01} that studied the properties of a model similar to ours (but without second-neighbor $J_2$ coupling along the chains) as a function of leg to diagonal coupling. 

In the other limit, $\eta \gg 1$, singlets on $J_1$ bonds are dominant. For $J_2=0$ the triplet sector of the nanotube model can be mapped onto the $S=1$ Haldane model~\cite{timonenContinuumlimitCorrelationFunctions1985,schulzPhaseDiagramsCorrelation1986,Scalapino00} and this limit thus corresponds to the Haldane phase~\cite{PhysRevLett.50.1153,PhysRevLett.59.799, Pollmann10, Pollmann12}. This phase exhibits a fourfold degenerate ground state for OBC due to localized and protected spin-1/2 edge excitations. One can notice this degeneracy in Fig.~\ref{fig:dmrg-op-and-gap}\hyperlink{fig:dmrg-op-and-gap}{(c)} and~\ref{fig:dmrg-op-and-gap}\hyperlink{fig:dmrg-op-and-gap}{(d)} for $\eta\ge0.65$ and $\eta\ge0.93$, respectively. The Haldane phase is a well-known example of a nontrivial symmetry-protected topological (SPT) phase~\cite{Pollmann10, Pollmann12}. SPTs are gapped phases with a symmetry protecting the topology of the state. In contrast, the RS phase is topologically trivial as it is adiabatically connected to a product state~\cite{Pollmann12}. In order to determine the topology of the $\mathbb{Z}_2$-helical phase, we investigate the degeneracy of the entanglement spectrum (ES) as a function of $\eta$ in Sec.~\ref{subsec:ES}. 

\subsection{Symmetries and symmetry breaking}
Let us now discuss in more detail how the $\mathbb{Z}_2$-helical phase transforms under the different symmetries of the nanotube model. This will also elucidate the relation to the previously found columnar dimer (CD) ordered phase~\cite{starykhDimerizedPhaseTransitions2004, liuExistenceDimerizedPhases2008, hikiharaPhaseDiagramFrustrated2010}, and demonstrate that the $\mathbb{Z}_2$-helical phase is different. We focus on the the $N_2 = 2$ nanotube model, which is depicted in Fig.~\ref{fig1trial}\hyperlink{fig1trial}{(b)}. but our analysis can easily be generalized to other even values of $N_x$. Note that Fig.~\ref{fig1trial}\hyperlink{fig1trial}{(c)} shows how the $N_x=2$ nanotube model continuously transforms to a two-leg spin ladder. We place the nanotube with its long axis along the $y$-axis, and choose the origin to lie at the center of the circle containing red sites. The unit cell vector $a_y$ connects two blue sites along the chain, and the unit cell thus contains four basis sites (two "blue" and two "red" sites). The nanotube model is invariant under the following set of generating symmetries
\begin{align}
    \mathcal{G} &= \{(\openone | 0), (\openone | 1),  (C_{4y}|\frac12), (C_{2y}|0), (C_{2x}|0), (\mathcal{P}|0), (S_4|0),  \nonumber \\ &\qquad (m_{h,b}|0), (m_{h,r}|0), (m_{v,b}|0), (m_{v,r}|0)\}\,.
    \label{eq:symmetries_G}
\end{align}
Here, a general element $(\mathcal{O}| t)$ combines a point group operation $\mathcal{O}$ with a translation $t$ along $y$, where $t=1$ for the translation between two adjacent unit cells. The element $\mathcal{P}$ denotes inversion around the center of a blue circular plaquette, $S_4$ includes a fourfold rotation and a mirror operation with mirror between blue and red sites at $a_y/4$. There are two horizontal mirror planes $m_{h,b}$ and $m_{h,r}$ that are parallel to the $xz$ plane and cut through the blue ($a_y = 0)$ and red sites ($a_y = 1/2$), respectively. Finally, there are two vertical mirror planes $m_{v,b}$ and $m_{v,r}$ that contain the $y$ axis and cut through blue or red sites, respectively. In the two-leg ladder description, the $(C_{4y}|\frac12)$ operation and the vertical mirror planes only exist if nearest-neighbor ($J_{nn}$) and diagonal bond ($J_\times$) couplings are equal (both are equal to $J_1$ in our model). Previous works~\cite{ vekuaQuantumDimerPhases2006, liuExistenceDimerizedPhases2008}, have not explored this parameter regime and focused on $J_{nn} \neq J_\times$, where $(C_{4y}|\frac12)$, $m_{v,b}$, and $m_{v,r}$ (and also $C_{2y}, C_{2x}, P, S_4$) are explicitly broken. 

\begin{figure}[tb]
    \centering
    \includegraphics[width=\linewidth]{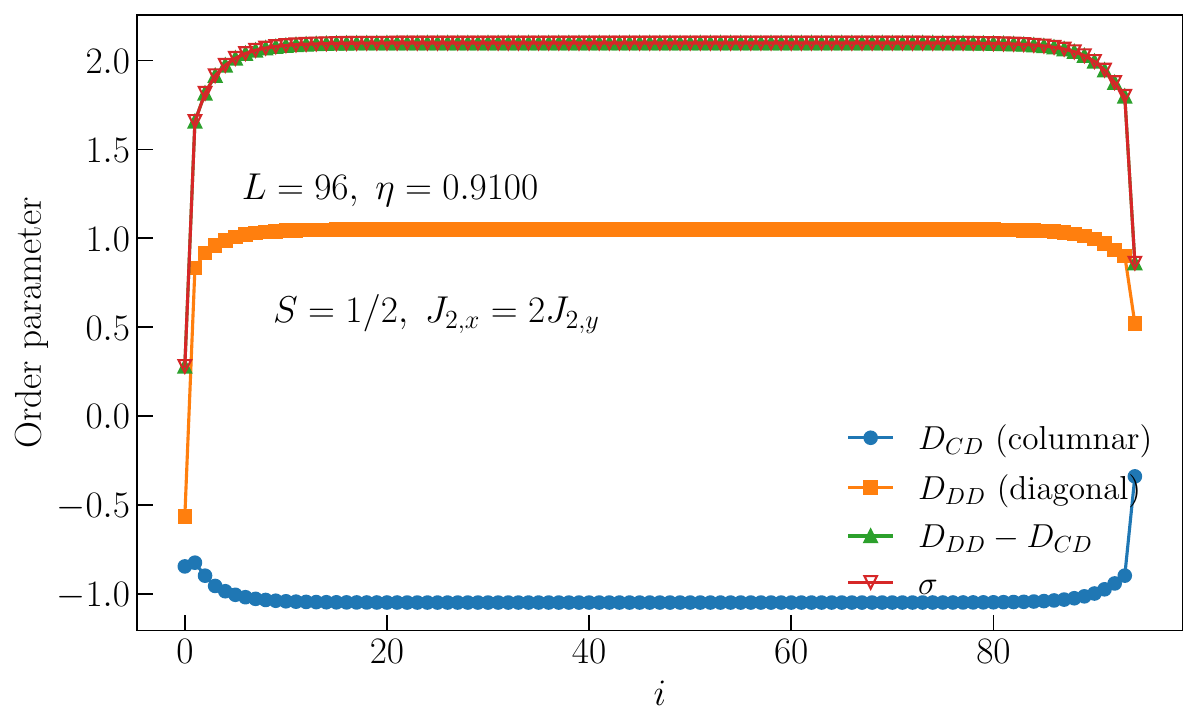}
    \caption{Local expectation value of the bond order parameters $D_{\text{CD}}$, $D_{\text{DD}}$, their difference, and $\sigma_i$ as a function of site $i$ in the $\mathbb{Z}_2$-helical phase at $\eta = 0.91$ in the nanotube geometry $J_{2,x} = 2 J_{2,y}$. The system size is set to $L=96$. One observes that the $\mathbb{Z}_2$-helical state exhibits $\langle D_{\text{DD},i} \rangle = - \langle D_{\text{CD},i} \rangle 1$ away from the edges and is thus distinct from pure CD or pure DD order (and also from a state with $D_{\text{DD}} = D_{\text{CD}}$).  }
    \label{fig:CD_DD_orders_versus_i}
\end{figure}

One can rewrite the $\mathbb{Z}_2$-helical bond order parameter $\sigma_i$ in Eq.~\eqref{eq:OP_sigma_i} as $\sigma_i = D_{\text{DD},i} - D_{\text{CD},i}$, where 
\begin{align}
    D_{\text{DD},i} &= \mathbf{S}_{4i-1} \cdot (\mathbf{S}_{4i-3} - \mathbf{S}_{4i+1}) + \mathbf{S}_{4i} \cdot (\mathbf{S}_{4i-2} - \mathbf{S}_{4i+2}) \\
    D_{\text{CD}, i} &= \mathbf{S}_{4i-1} \cdot (\mathbf{S}_{4i-2} - \mathbf{S}_{4i+2})
    + \mathbf{S}_{4i} \cdot (\mathbf{S}_{4i-3} - \mathbf{S}_{4i+1}) 
\end{align}
are the local bond order parameters of diagonal dimer (DD) and columnar dimer (CD) states, respectively. The CD phase is characterized by a nonzero ground state expectation value of $\frac{1}{\mathcal{N}} \sum_{i=1}^{L_y-1} 
D_{\text{CD},i}$~\cite{hikiharaPhaseDiagramFrustrated2010}, and analogously for the DD phase. 
One can readily observe that out of the set of symmetries $\mathcal{G}$ in Eq.~\eqref{eq:symmetries_G}, the elements $\{\mathcal{P}, S_4, m_{h,b}, m_{h,r}, m_{v,b}, m_{vr}\}$ are spontaneously broken in the $\mathbb{Z}_2$-helical phase. Importantly, however, the $\mathbb{Z}_2$-helical phase preserves $(C_{4y}|\frac12)$ symmetry. In contrast, the application of $(C_{4y}|\frac12)$ transforms a state with homogeneous CD order $\langle D_{\text{CD},i} \rangle = 
D_{\text{CD}}$ into a state with homogeneous DD order $\langle D_{\text{DD},i} \rangle = D_{\text{DD}}$, where $D_{\text{CD}} = - D_{\text{DD}}$. The breaking of $(C_{4y}|\frac12)$ in a state with pure CD or pure DD order thus distinguishes these orders from $\mathbb{Z}_2$-helical order. Note that a state with both CD and DD order present but $D_{\text{CD}} = + D_{\text{DD}}$ (such that $\sigma = 0$) also breaks $(C_{4y}|\frac12)$ symmetry. While dimers are aligned either along the chain (for CD order) or along the diagonals (for DD order), the $\mathbb{Z}_2$-helical order is characterized by resonating singlets between configurations that fulfill $D_{\text{CD},i} = -D_{\text{DD},i}$. In Fig.~\ref{fig:CD_DD_orders_versus_i} we show the local expectation value of $D_{\text{CD},i}$ and $D_{\text{DD},i}$ across the chain in the $\mathbb{Z}_2$-helical phase, which indeed exhibits $\langle D_{\text{CD},i} \rangle = - \langle D_{\text{DD},i} \rangle$ in the bulk. 
\begin{figure*}[tbh]
    \centering
    \includegraphics[width=\linewidth]{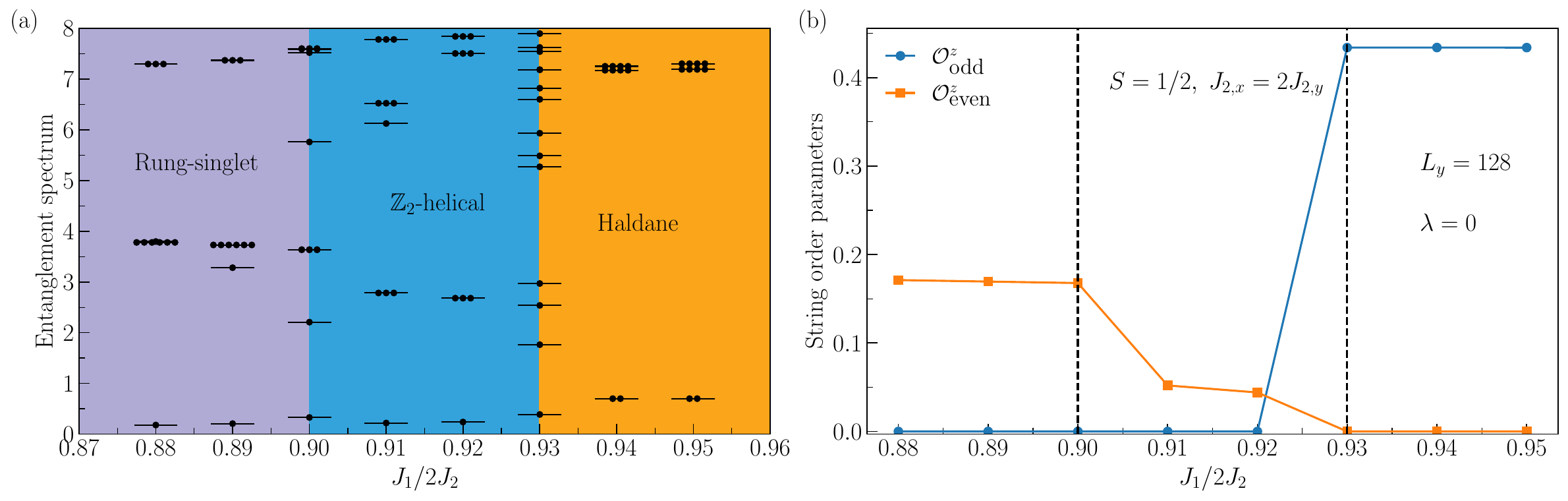}
    \caption{(a) Entanglement spectrum and (b) string order parameters $\mathcal{O}^z_\textrm{odd}$ and $\mathcal{O}^z_\textrm{even}$ as a function of $\eta = J_1/2 J_2$ for a nanotube with $J_{2x} = 2J_{2y}$ and length $L_y=256$ and no pinning field ($\lambda=0$). The dashed line in panel (b) denotes the $\mathbb{Z}_2$-helical phase.}
    \label{fig:dmrg-entanglement}
\end{figure*}

\subsection{Entanglement spectrum and string order parameters}
\label{subsec:ES}
In order to show that the two phases surrounding the $\mathbb{Z}_2$-helical phase are indeed the topologically trivial RS and topologically nontrivial Haldane phases, we measure the entanglement spectrum for a cylinder of length $L_y=256$ for the case $J_{2x} = 2J_{2y}$ and without pinning field ($\lambda=0$). It has been shown that the lowest entanglement level for the RS and the Haldane phase are singly and doubly degenerate, respectively~\cite{Pollmann10,Pollmann12}. In Fig.~\ref{fig:dmrg-entanglement}\hyperlink{fig:dmrg-entanglement}{(a)} one can notice the same degeneracy of the lowest level as expected for these two phases. The $\mathbb{Z}_2$-helical phase displays degeneracies with the same parity as the RS phase, thus suggesting that it is not a nontrivial SPT phase.

\begin{figure*}[tbh]
    \centering
    \includegraphics[width=\linewidth]{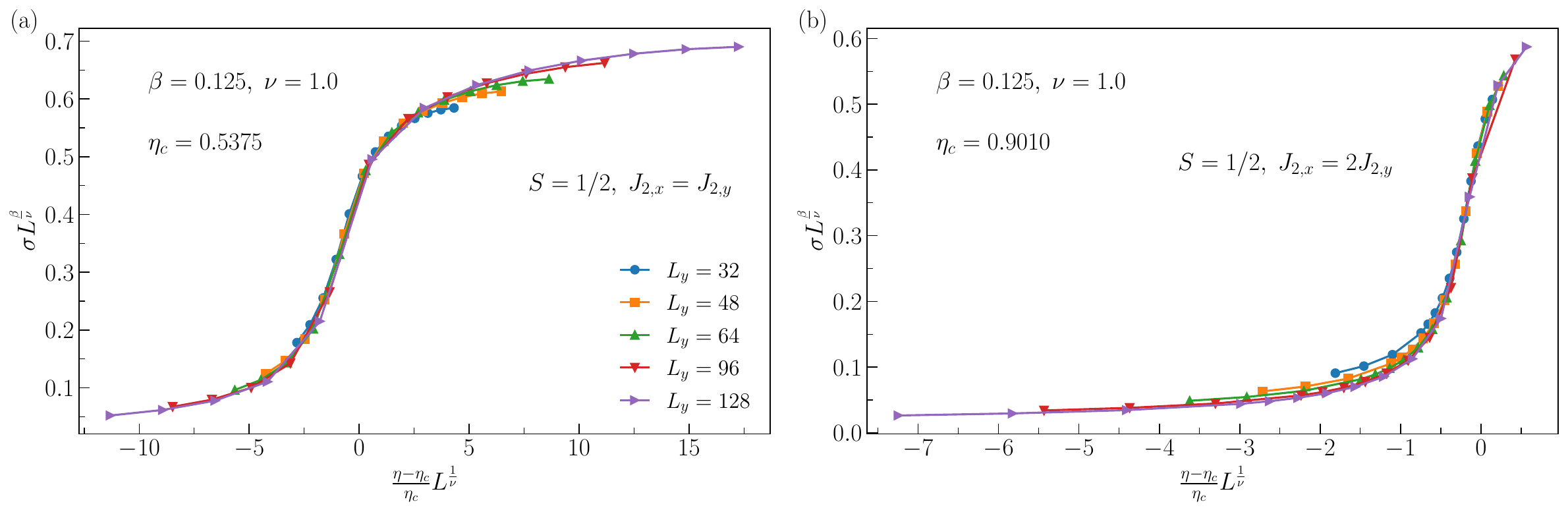}
    \caption{Finite-size scaling data collapse of the $\mathbb{Z}_2$-helical order parameter $\sigma$ for (a) nanostrip $J_{2x} = J_{2y}$ and (b) nanotube $J_{2x} = 2 J_{2y}$ geometries. The critical exponents are set to be those of the 2D Ising model universality class, i.e., $\beta = 1/8$ and $\nu = 1$. The location of the critical point $\eta_c \equiv \eta_{c,1}$ is set to the values shown in the figure. }
    \label{fig:dmrg-sigma-collapse}
\end{figure*}

Another way of characterizing the RS and Haldane phases is by measuring certain nonlocal string order parameters. We consider the same definition as given in Ref.~\cite{Solyom01} for the odd and even string order parameters: the former pairing nearest neighbors across rungs, whereas the latter pairs diagonal nearest neighbors. The notation of ``even" and ``odd" order parameters arises from the parity of the number of singlet bonds that are broken if one horizontally cuts the spin ladders depicted in Fig.~\ref{fig:ops}\hyperlink{fig:ops}{(b,~c)}~\cite{Solyom01}.  In order to minimize boundary effects, we select pairs formed by sites $\ell=N/4$ and $N-\ell$, where $N=4L_y$ is the total number of sites, as the starting and end points of the string order parameters, respectively. Therefore, given the site labelling of Fig.~\ref{fig:ops}\hyperlink{fig:ops}{(a)}, the equations for the string order parameters are
\begin{align}\mathcal{O}^z_\textrm{odd} = -\left(S^z_{\ell} + S^z_{\ell-1}\right) & \prod_{j=\ell+1}^{N-\ell-2} \exp\left(i \pi S^z_j\right) \times\nonumber\\
    & \left(S^z_{N-\ell} + S^z_{N-\ell-1}\right),
    \label{eq:odd_string_op}
\end{align}
for the odd one, and
\begin{align}
    \mathcal{O}^z_\textrm{even} = -\left(S^z_{\ell} + S^z_{\ell+2}\right) & e^{i \pi S^z_{\ell+1}} \prod_{j=\ell+3}^{N-\ell-1}\exp \left(i\pi S^z_j\right) \times\nonumber\\
    & \left(S^z_{N-\ell} + S^z_{N-\ell+2}\right),
    \label{eq:even_string_op}
\end{align}
for the even. Here and in the following, we assume $L_y$ to be even. 
Hence, according to these definitions, $\mathcal{O}^z_\textrm{odd}\rightarrow0$ if the ground state has predominantly rung singlets (RS phase), whereas $\mathcal{O}^z_\textrm{even}\rightarrow0$ when the ground state has a majority of diagonal singlets (Haldane phase). 

In Fig.~\ref{fig:dmrg-entanglement}\hyperlink{fig:dmrg-entanglement}{(b)} we show the two string order parameters, $\mathcal{O}^z_\textrm{odd}$ and $\mathcal{O}^z_\textrm{even}$, as a function of $\eta$ for a nanotube of length $L_y=256$. We focus on the nanotube case $J_{2x} = 2 J_{2y}$, but we expect the result to be the same for the nanostrip. As we show in Appendix~\ref{appendix:scaling}, the data has already saturated for the chosen cylinder length $L_y$ and these results are thus representative of the thermodynamic limit. Notice that $\mathcal{O}^z_\textrm{odd}\ne0$ only in the Haldane phase, as expected. On the other hand, $\mathcal{O}^z_\textrm{even}$ remains finite within both the RS and the $\mathbb{Z}_2$-helical phase. Its value in the $\mathbb{Z}_2$-helical phase is about four times less than the value of $\mathcal{O}^z_\textrm{even}$ at the RS phase.

\subsection{Ising criticality}
\label{subsec:Ising_criticality}
Finally, after having established the three different phases within the $J_1-J_2$ nanotube model, let us return to the characterization of the critical point and the universality class of the transition at $\eta_{c,1}$. Using the results for the  order parameter in the upper panels of Fig.~\ref{fig:dmrg-op-and-gap}, we  perform a finite-size scaling analysis, which yields an estimate for the critical exponents $\beta$ and $\nu$. In Fig.~\ref{fig:dmrg-sigma-collapse} we show the outcome of the scaling analysis adopting the 2D Ising universality class critical parameters $\beta=1/8$ and $\nu=1$. We find a satisfactory collapse of the data points for both nanostrip and nanotube geometries.

\begin{figure}[t!]
    \centering
    \includegraphics[width=\linewidth]{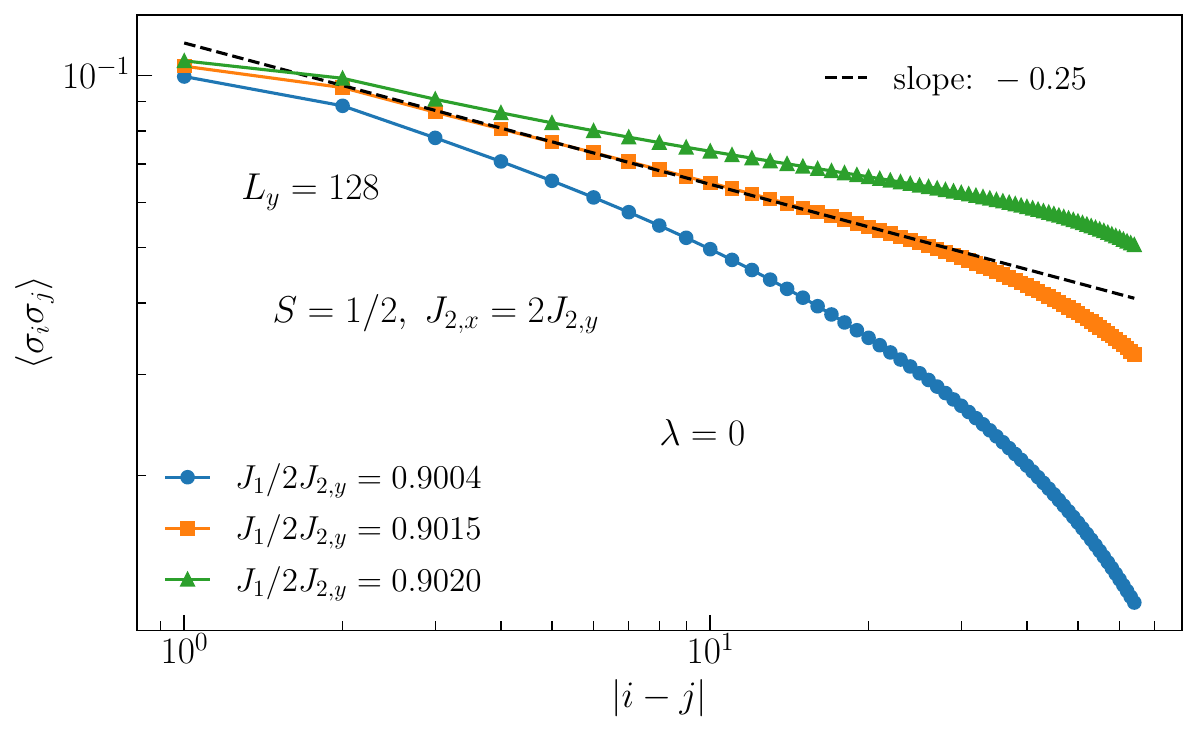}
    \caption{Correlation function of the order parameter $\left\langle \sigma_i \sigma_j \right\rangle$ versus the separation between sites $\left|i-j\right|$ on a log-log scale for various $\eta = J_1/2 J_2$ values. The data correspond to a cylinder of length $L_y=128$ for the nanotube case $J_{2x} = 2 J_{2y}$. The symmetry-breaking pinning field is set to zero, $\lambda = 0$. We find that the correlations decay algebraically with the expected 2D Ising exponent $\eta = 0.25$ for $J_1/(2 J_{2y}) = 0.9015$. The algebraic decay is observed up to a separation of about $|i-j| \approx 30$ when finite size effects set in. For $J_1/(2J_{2y})$ smaller than the critical value, the correlations decay exponentially, and they tend to saturate at a nonzero value for $J_1/(2J_{2y})> 0.9015$. }
    \label{fig:dmrg-opcorr}
\end{figure}

One can obtain another critical exponent by directly evaluating the correlation function of the order parameter $\left\langle \sigma_i \sigma_j \right\rangle$. In order to reduce the effects of the open boundaries, we start from the plaquettes $i$ and $j$ in the middle of the cylinder and increase the separation between plaquette pairs by moving towards both ends of the nanotube. In Fig.~\ref{fig:dmrg-opcorr} we show $\left\langle \sigma_i \sigma_j \right\rangle$ as a function of $\left| i - j \right|$ for a nanotube of length $L_y=128$ and three different $\eta$ values close to the critical point. Notice that we find an excellent agreement with the expected critical exponent value of the 2D Ising universality class for $\eta=0.9015$, as shown by the power-law fitted dashed line.

Finally, we comment on the the cost of the DMRG simulations for the nanostrip versus the nanotube setups. We observed that the nanostrip case $J_{2x} = J_{2y}$ requires a significantly larger bond dimension than the nanotube case to achieve the same accuracy. For instance, for a cylinder of length $L_y=64$ and choosing $\eta=0.64$ and $\eta=0.92$ for nanostrip and nanotube, respectively, inside the  $\mathbb{Z}_2$-helical phase near the first-order phase transition, the DMRG run for the nanostrip reaches the desired accuracy with a maximum bond dimension $m=1322$, whereas it requires only $m=414$ for the nanotube. This indicates that the ground state in the nanostrip case may be more complex and thus harder to converge to, which may deserve further studies. 

\section{Discussion}
\label{sec:discussion}
In summary, both analytic and computational methods indicate that the $\mathbb{Z}_2$ nematic order of the two-dimensional $J_1-J_2$ latice model survives in the (1+1)-dimensional quantum limit of spin nanotubes.  Furthermore the disordered states on either side of the $\mathbb{Z}_2$ ordered phase are distinct with different string order parameters, a feature that emerges already in the Schwinger boson
treatment; from the two-dimensional perspective these distinct
spin liquids were not anticipated, though they are qualitatively consistent with results for two-leg ladders, equivalent to our $N_x=2$ spin nanotubes. We should note that we are not aware of any ladder calculations performed with the frustrated coupling profile that we present here so a detailed comparison is not possible.

We have focused on gapped spin nanotubes with even $N_x$ legs.  The Schwinger Boson (SB) mean-field theory predicts two short-range spin liquid phases at small and large $J_1/J_2$ that are separated by an emergent $\mathbb{Z}_2$ helical phase for all values of $S$. This result is confirmed computationally for $S=1/2$ and $N_x=2$ systems using Density Matrix Renormalization Group (DMRG).  However, the DMRG results indicate that the $\mathbb{Z}_2$-helical phase is narrower than what was predicted by SB theory, though it is still present.  Nevertheless, both methods agree that the
$\mathbb{Z}_2$-helical phase is larger for the nanostrip geometry $(J_{2x} = J_{2y})$ compared to the nanotube geometry ($J_{2x} = 2 J_{2y})$.

The $\mathbb{Z}_2$-helical phase is characterized by a local order parameter.  Adopting the language of the ladder literature, it lies between RS and Haldane spin liquid phases that are identified by distinct nonlocal even and odd string order parameters, which are defined in Fig.~\ref{fig:ops}\hyperlink{fig:ops}{(b, c)}. This raised the interesting question of whether the local order parameter in the $\mathbb{Z}_2$-helical phase coexists with any of the nonlocal string orders and whether the phase is a topologically nontrivial SPT phase. We find that the even string order parameter of the RS phase is reduced but remains nonzero in the $\mathbb{Z}_2$-helical phase. By computing the entanglement spectrum and finding that it does not exhibit any degeneracies, however, we conclude that the $\mathbb{Z}_2$-helical phase is topologically trivial, just like the nearby RS phase. It interpolates between the RS and nontrivial SPT Haldane phase, where the latter exhibits a twofold degenerate entanglement spectrum and an odd string order parameter that vanishes in the RS and $\mathbb{Z}_2$-helical phases.

There are many questions that may motivate future research. 
First, we have here focused on gapped spin nanotubes with an even number $N_x$ of spins along the short direction. Investigating half-integer spin chains with odd $N_x$, which are gapless, opens the possibility to explore the emergence of discrete symmetry breaking in models of algebraically correlated spins. Second, DMRG studies for higher spins $S > 1/2$ chains might explore the possible absence of the Haldane phase for integer spin. Third, the identification and scaling of the energy operator as well as other operators of the underlying conformal field theory~\cite{zouConformalFieldsOperator2020} would confirm the 2D Ising nature of the quantum critical point. Similarly, the scaling of the entanglement entropy with the bond dimension could be used to extract the central charge of the theory. Building a multi-scale entanglement renormalization ansatz (MERA) of the state at the quantum critical point would allow for further investigation of this emergent phase transition driven by the condensation of a composite spin order parameter~\cite{vidalEntanglementRenormalization2007,arguello-luengoGeneralizedMultiscaleEntanglement2022}.

Finally, these spin nanotube models could be studied using bosonization~\cite{Tsvelik96, allenNonAbelianBosonizationFrustrated1997, vekuaQuantumDimerPhases2006}. The $J_1=0$ limit of our $N=2$ model corresponds to two decoupled $S=1/2$ spin ladders; from previous bosonization studies~\cite{Tsvelik96}, we know  that the low energy spectrum of these ladders involves massive $S=0$ singlet and $S=1$ triplet Majorana excitations. We expect this description to adiabatically evolve as  $J_1$ becomes finite and the ladders are coupled, which immediately  suggests that the $\mathbb{Z}_2$ transition 
can be described within a Majorana framework~\cite{allenNonAbelianBosonizationFrustrated1997,vekuaQuantumDimerPhases2006}. Since the $\mathbb{Z}_2$ transition does not involve any broken spin symmetry, the gapped triplet Majorana degrees of freedom will decouple from the transition, as supported by our computational studies. By contrast, the two singlet Majoranas together will form a 
gapped complex fermionic excitation whose mass will vanish and change sign at the transition, reminiscent of
the transverse field Ising model~\cite{vekuaQuantumDimerPhases2006}. These ideas motivate future bosonization studies, exploring the interplay of local order parameters, symmetry breaking and symmetry-protected topological order.

\acknowledgements
We thank Elio Koenig for discussions related to this work. We also appreciate the referee's suggestion of a possible Majorana description of the $\mathbb{Z}_2$ phase transition described here. 
This work was supported by the U.S. Department of Energy (DOE), Office of Science, Basic Energy Sciences, Division of Materials Sciences and Engineering, including the grant of computer time at the National Energy Research Scientific Computing Center (NERSC) in Berkeley, California (J.C.G. and P.P.O), by DOE Basic Energy Sciences grant DE-SC0020353 (P.Ch. and Z.Z.) and NSF grant DMR-1830707 (P. Co.). Part of the research (J.C.G. and P.P.O.) was performed at the Ames National Laboratory, which is operated for the U.S. DOE by Iowa State University under Contract DE-AC02-07CH11358. This work was supported by a Leverhulme Trust International Professorship grant number LIP-202-014 (S.L.S). For the purpose of Open Access, the author has applied a  CC  BY  public  copyright  licence  to  any Author  Accepted  Manuscript  version  arising  from  this submission. P. Chandra thanks Princeton University for hospitality when this project was initiated. P.P.O. thanks A. Vishwanath and Harvard University for hospitality during the final stages of this project. 

\appendix
\section{Schwinger boson theory calculations}\label{appendix:SchBoson}
\subsection{Mean-field equations}
When $\eta\ll 1$ and $S$ is large, one expects that there are two interpenetrating short-range-ordered antiferromagnetic sublattices, which are locked together and form a collinear state due to the order by disorder effect. In the collinear state, we take the translationally invariant mean-field ansatz shown in Fig. \ref{fig:SB_ansaetze} (denoted as $\mathbb{Z}_2$ ansatz), which is consistent with the correlation of the corresponding state. When $S$ decreases, the quantum fluctuation melts the collinear order and the two sublattices become decoupled. In this state, the expectation values $\Delta_+$ and $h_-$ should vanish simultaneously so that there is no correlation between the two decoupled antiferromagnetic sublattices (DAS) \cite{Flint2009}. With this ansatz, the mean field Hamiltonian can be diagonalized by Fourier transform and Bogoliubov transformation. 

For $N_x\geq 4$, the ground state energy per Kramers pair (or $N=2$) per site is given by
\begin{multline}
    \frac{E_{\mathbb{Z}_2}}{N_s}=\frac{4}{J_1}\Delta_+^2-\frac{4}{J_1}h_-^2+\frac{4}{J_2}\Delta_x^2+\frac{4}{J_2}\Delta_y^2-\lambda(2S+1)\\+\frac{1}{N_x}\sum_{k_x}\int_{-2\pi}^{2\pi}\frac{dk_y}{4\pi}\omega_{\mathbb{Z}_2}(k_x,k_y), \label{Z2Energy}
\end{multline}
where the bosonic spectrum reads
\begin{equation}
    \omega_{\mathbb{Z}_2}=\sqrt{s^2-t^2},
\end{equation}
and
\begin{equation}
    s=\lambda+2h_-\cos{\frac{k_x-k_y}{2}},
\end{equation}
\begin{equation}
    t=2\left(\sin{\frac{k_x+k_y}{2}}\Delta_+ +\sin{k_x}\Delta_x+\sin{k_y}\Delta_y\right).
\end{equation}
\begin{figure}[tbh]
    \centering
    \includegraphics[width=0.3\textwidth]{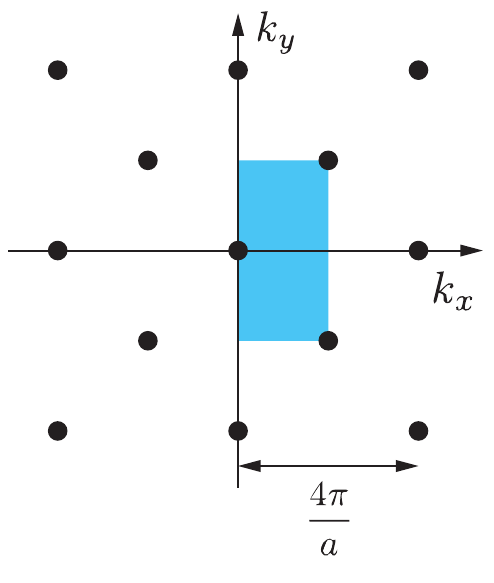}
    \caption{The choice of Brillouin zone (marked by blue region) in the Schwinger boson approach.}
    \label{fig:BZ}
\end{figure}
For $N_x=2$ case, the $J_2$ associated with $\Delta_y (\Delta_x)$ in Eq. (\ref{Z2Energy}) should be replaced by $J_{2y}(J_{2x}/2)$ while the rest of the equation remains unchanged (see our definition in Sec. \ref{sec:dmrg_results}). In the below, we only show equations derived for $N_x=4$ case and their modification for $N_x=2$ are straightforward. For convenience we have set lattice constant $a=1$ (see Fig. \ref{fig:SB_ansaetze}) and choose the Brillouin zone to be $0 \leq k_x<2\pi, -2\pi \leq k_y<2\pi$ (see Fig. \ref{fig:BZ}). As will be discussed later, in the $\mathbb{Z}_2$ ansatz $k_x$ takes discrete value $k_x=2\pi m/N_x+\pi/2$,  where $m=0,1,..,N_x-1$. By minimizing the ground state energy one obtains the set of mean field equations:
\begin{equation}
    2S+1=\frac{1}{N_x}\sum_{k_x}\int_{-2\pi}^{2\pi}\frac{dk_y}{4\pi} \frac{s}{\omega_{\mathbb{Z}_2}}, \label{MFS}
\end{equation}
\begin{equation}
   \frac{4}{J_1}h_-=\frac{1}{N_x}\sum_{k_x}\int_{-2\pi}^{2\pi}\frac{dk_y}{4\pi} \frac{s}{\omega_{\mathbb{Z}_2}}\cos\frac{k_x-k_y}{2}, \label{MFh}
\end{equation}
\begin{equation}
   \frac{4}{J_1}\Delta_+=\frac{1}{N_x}\sum_{k_x}\int_{-2\pi}^{2\pi}\frac{dk_y}{4\pi} \frac{t}{\omega_{\mathbb{Z}_2}}\sin\frac{k_x+k_y}{2}, \label{MFDeltap}
\end{equation}
\begin{equation}
   \frac{4}{J_2}\Delta_x=\frac{1}{N_x}\sum_{k_x}\int_{-2\pi}^{2\pi}\frac{dk_y}{4\pi} \frac{t}{\omega_{\mathbb{Z}_2}}\sin{k_x}, \label{MFDeltax}
\end{equation}
\begin{equation}
   \frac{4}{J_2}\Delta_y=\frac{1}{N_x}\sum_{k_x}\int_{-2\pi}^{2\pi}\frac{dk_y}{4\pi} \frac{t}{\omega_{\mathbb{Z}_2}}\sin{k_y}. \label{MFDeltay}
\end{equation}
To determine whether a continuous phase transition occurs, we also need the Hessian matrix elements $Q=\partial^2 E_{\mathbb{Z}_2} /(N_s\partial \alpha_i\partial\alpha_j)$, where 
\begin{equation}
\vec{\alpha}=\left(\lambda,\Delta_x,\Delta_y,h_-,\Delta_+\right)
\end{equation}
is the shorthand for the mean-field parameters. It turns out that when $h_-=\Delta_+=0$ the $5\times 5$ Hessian matrix $Q$ is block diagonal
\begin{equation}
    Q=\begin{pmatrix}
    X & 0\\
    0 & Y\\
    \end{pmatrix}
\end{equation}
where
\begin{widetext}
\begin{equation}
    X=\begin{pmatrix}
        -\bigintss_{k}\dfrac{t^2}{\omega^3} &2\bigintss_{k}\dfrac{\sin k_x st}{\omega^3}& 2\bigintss_{k}\dfrac{\sin k_y st}{\omega^3} \\[2ex]
        2\bigintss_{k}\dfrac{\sin k_x st}{\omega^3} & \dfrac{8}{J_2}-4\bigintss_{k}\dfrac{\sin^2 k_x s^2}{\omega^3} &-4\bigintss_{k}\dfrac{\sin k_x\sin k_y s^2}{\omega^3}\\[2ex]
        2\bigintss_{k}\dfrac{\sin k_y st}{\omega^3} & -4\bigintss_{k}\dfrac{\sin k_x\sin k_y s^2}{\omega^3}&\dfrac{8}{J_2}-4\bigintss_{k}\dfrac{\sin^2 k_y s^2}{\omega^3}\\[2ex]
    \end{pmatrix},
\end{equation}
\begin{equation}
    Y=\begin{pmatrix}
       -\dfrac{8}{J_1}-  4\bigints_{k}\dfrac{ \cos^2 \frac{k_x-k_y}{2}t^2}{\omega^3}& 4\bigints_{k} \dfrac{\cos \frac{k_x-k_y}{2}\sin \frac{k_x+k_y}{2} st}{\omega^3}\\[2ex]
        4\bigints_{k} \dfrac{\cos \frac{k_x-k_y}{2}\sin \frac{k_x+k_y}{2} st}{\omega^3}& \dfrac{8}{J_1}-4\bigints_{k}\dfrac{\sin^2 \frac{k_x+k_y}{2} s^2}{\omega^3}\\[2ex]
    \end{pmatrix}\label{Hessian45}
\end{equation}
\end{widetext}
and we have used shorthand $\int_{k}\equiv\frac{1}{N_x}\sum_{k_x}\int_{-2\pi}^{2\pi}\frac{dk_y}{4\pi}$, $\omega\equiv\omega_{\mathbb{Z}_2}$.

When $\eta\gg 1$, the system is expected to have N\'eel short range order (SRO). In the symplectic-N Schwinger boson theory, we consider the mean-field ansatz (N\'eel ansatz) shown in Fig. \ref{fig:SB_ansaetze}. The ground state energy per site can be shown to be 
\begin{multline}
    \frac{E_N}{N_s}=\frac{4}{J_1} \Delta_{+}^2+\frac{4}{J_1} \Delta_{-}^2-\frac{4}{J_2}h_x^2-\frac{4}{J_2}h_y^2-\lambda(2S+1)\\
    +\frac{1}{N_x}\sum_{k_x}\int_{-2\pi}^{2\pi}\frac{dk_y}{4\pi}\omega_N(k_x,k_y), \label{NeelEnergy}
\end{multline}
where $k_x$ takes discrete values $k_x=2\pi m/N_x$, $m=0,1,...,N_x-1$. The bosonic spectrum is 
\begin{equation}
    \omega_N=\sqrt{u^2-v^2},
\end{equation}
where
\begin{equation}
    u=\lambda+2h_x \cos k_x+2h_y\cos k_y,
\end{equation}
and
\begin{equation}
    v=2\sin\frac{k_x-k_y}{2}\Delta_- -2\sin\frac{k_x+k_y}{2}\Delta_+.
\end{equation}
The self-consistent mean-field equations are 
\begin{equation}
    2S+1=\frac{1}{N_x}\sum_{k_x}\int_{-2\pi}^{2\pi}\frac{dk_y}{4\pi} \frac{u}{\omega_N}, 
\end{equation}
\begin{equation}
   \frac{4}{J_1}\Delta_-=\frac{1}{N_x}\sum_{k_x}\int_{-2\pi}^{2\pi}\frac{dk_y}{4\pi} \frac{v}{\omega_N}\sin\frac{k_x-k_y}{2}, 
\end{equation}
\begin{equation}
   \frac{4}{J_1}\Delta_+=-\frac{1}{N_x}\sum_{k_x}\int_{-2\pi}^{2\pi}\frac{dk_y}{4\pi} \frac{v}{\omega_N}\sin\frac{k_x+k_y}{2}, 
\end{equation}
\begin{equation}
   \frac{4}{J_2}h_x=\frac{1}{N_x}\sum_{k_x}\int_{-2\pi}^{2\pi}\frac{dk_y}{4\pi} \frac{u}{\omega_N}\cos{k_x}, 
\end{equation}
\begin{equation}
   \frac{4}{J_2}h_y=\frac{1}{N_x}\sum_{k_x}\int_{-2\pi}^{2\pi}\frac{dk_y}{4\pi} \frac{u}{\omega_N}\cos{k_y}.
\end{equation}
 By solving the above equations self-consistently and comparing the energies Eqs. (\ref{Z2Energy}) and (\ref{NeelEnergy}), one can determine the first-order phase transition lines shown in the main text figures.
\begin{figure}
    \centering
    \includegraphics[width=0.48\textwidth]{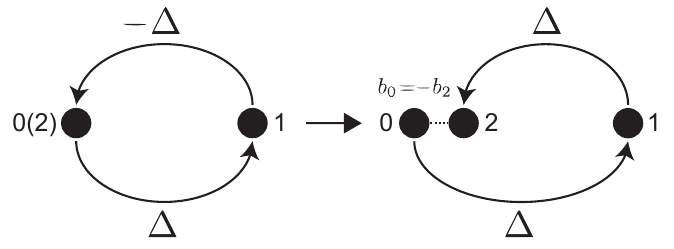}
    \caption{The mean-field ansatz of a two-site Heisenberg model. Left: Boson wavefunction is periodic and the mean-field ansatz is not translationally invariant along the loop. Right: Wavefunction is anti-periodic but the ansatz has translational symmetry.}
    \label{fig:Nx2}
\end{figure}
\subsection{$\mathbb{Z}_2$ phase transition}
In this section, we use Schwinger boson theory to analytically investigate the $\mathbb{Z}_2$ phase transition of spin nanotube when $\eta \ll 1$.
\subsubsection{$N_x=2$ case}

We first consider the case (ii) $J_{2x}=2J_{2y}=2J$, for which we refer as the spin nanotube. In the $N_x=2$ case, the bosons have anti-periodic instead of periodic boundary condition along the $x$-direction, so $k_x$ takes value $ \frac{\pi}{2},\frac{3\pi}{2}$. To see this, one can consider a simpler case: a two-site antiferromagnetic model on a ring shown in Fig. \ref{fig:Nx2}. Because $\Delta_{12}=-\Delta_{21}$, to make the bosonic Hamiltonian translationally invariant with constant $\Delta$ around the ring, one has to apply a singular gauge transformation leaving $\Delta_{10}, b_0, b_1$ invariant but $b_{2}\rightarrow -b_2, \Delta_{12}\rightarrow -\Delta_{12}$, which makes the wavefunction anti-periodic. When $h_-=\Delta_+=0$ Eqs. (\ref{MFS})-(\ref{MFDeltay}) can be simplified to
\begin{equation}
    2S+1=\int_{-\pi}^{\pi}\frac{dk_y^\prime}{2\pi}\frac{1}{\sqrt{1-z^2}}, \label{MFSNx2}
\end{equation}
\begin{equation}
    \frac{4}{J_2}\Delta_x=\int_{-\pi}^{\pi}\frac{dk_y^\prime}{2\pi}\frac{z}{\sqrt{1-z^2}}, \label{MFDeltaxNx2}
\end{equation}
\begin{equation}
    \frac{4}{J_2}\Delta_y=\int_{-\pi}^{\pi}\frac{dk_y^\prime}{2\pi}\frac{z\cos k_y^\prime}{\sqrt{1-z^2}}, \label{MFDeltayNx2}
\end{equation}
where for convenience we have made substitution $k_y\rightarrow k_y^\prime+\frac{\pi}{2}$ and $z=2(\Delta_x+\cos{k_y}\Delta_y)/\lambda$. 

When $S$ is small, the only two non-zero mean-field parameters are $\lambda,\Delta_x$ or $\lambda,\Delta_y$, and we verify that the former one is energetically favorable. Solving Eq.(\ref{MFSNx2}) and Eq. (\ref{MFDeltaxNx2}) one obtains
\begin{equation}
    \lambda=J_2(S+\frac{1}{2}), \quad\Delta_x=\frac{J_2\sqrt{S(S+1)}}{2}, \label{LowSsol}
\end{equation}
which means at small $S$, the ground state is a valence bond solid in which every two sites connected by the $J_2$ bond in the $x$-direction form singlets. This is because we are considering the case $J_{2x}=2J_{2y}$ where the $J_2$ coupling along the $x$-direction is doubled and is twice as large as the one along the $y$-direction, so singlet state is favored due to large quantum fluctuation. When $S$ further increases and approaches a lower critical value $S_\text{cl}$, the system becomes unstable to being antiferromagnetically short-range ordered, and $\Delta_y$ acquires a non-vanishing value. To see this, one could evaluate the Hessian matrix element $\frac{\partial^2 E_g}{\partial \alpha_3^2}$ (note that when $\Delta_y=h_-=\Delta_+=0$ the Hessian matrix elements $\frac{\partial^2 E_g}{\partial \alpha_3\partial\alpha_j}$ all vanish for $j\neq 3$), which vanishes at the transition point 
\begin{equation}
    \dfrac{8}{J_2}-4\int_{k}\dfrac{\sin^2 k_y s^2}{\omega_{\mathbb{Z}_2}^3}=0.
\end{equation}
With Eq.(\ref{LowSsol}), one obtains $S_\text{cl}=1/\sqrt{2}-1/2\approx 0.21$.

As $S$ further increases, if $J_1$ coupling is absent, $\lambda, \Delta_x, \Delta_y$ all increase and at large $S$ approaches $\Delta_x/\lambda\sim\Delta_y/\lambda\sim 1/4$. In this regime it is useful to define a small parameter $\epsilon=1-2\frac{\Delta_x+\Delta_y}{\lambda}$ and $z=(1-\epsilon)\cos^2 \frac{k_y^\prime}{2}+O(\frac{1}{\ln \epsilon})$ when $\epsilon\ll 1$. From Eqs. (\ref{MFSNx2})(\ref{MFDeltaxNx2}) and (\ref{MFDeltayNx2}) one obtains
\begin{equation}
    2S+1\approx\frac{1}{\sqrt{2}\pi}[5\ln{2}-\ln{\epsilon}],\quad \lambda\approx -\frac{J_2}{\sqrt{2}\pi}\ln{\epsilon}. \label{SvsEpsilon}
\end{equation} Note there are four degenerate quasi-gapless modes at $\vec{k}=(\pi/2,\pi/2), (\pi/2,-3\pi/2), (3\pi/2,3\pi/2), (3\pi/2,-\pi/2)$ with corresponding correlation length $\xi\sim 1/\sqrt{\epsilon}$, which are reminiscent of the Goldstone modes of two uncoupled antiferromagnets. When $J_1$ is present, at some point the system is unstable to the collinear order and $\Delta_+, h_-$ develops, which hybridizes the four degenerate quasi-gapless modes and gives rise to only two quasi-gapless modes. At the transition point, the determinant of the Hessian in the $(h_-,\Delta_+)$ sector
\begin{multline}
     \text{det}(Y)=\left(A_1+A_2+2B\right)\left(A_1+A_2-2B\right)\\-\left(A_1-A_2-\frac{8}{J_1}\right)^2, \label{Hessian} 
\end{multline}
should vanish. In the above equation we define $A_1=-2\int_k (\cos^2 \frac{k_x-k_y}{2}t^2)/\omega^3$, $A_2=-2\int_k (\sin^2 \frac{k_x+k_y}{2}s^2)/\omega^3$ and $B=2\int_k (\cos \frac{k_x-k_y}{2}\sin \frac{k_x+k_y}{2} st)/\omega^3$. One can show that $|A_1|,|A_2|,|B|$ are identical to all divergent orders, and the lowest order contribution is given by
\begin{equation}
    A_1\sim A_2\sim -B\approx \frac{2}{J_2\epsilon\ln{\epsilon}}. \label{A1}
\end{equation}
Nevertheless, their differences are convergent in the limit $\epsilon\rightarrow 0$ and particularly,
\begin{multline}
    A_1+A_2+2B=\frac{1}{2\pi\lambda}\int_{-\pi}^{\pi}dk_y^\prime\frac{(z-1)(1+\cos{k_y^\prime})}{(z+1)\sqrt{1-z^2}}\\=-\frac{\gamma_2}{2\sqrt{2}\pi\lambda}\approx\frac{\gamma_2}{2J_2\ln \epsilon}, \label{A1A2B}
\end{multline}
\begin{equation}
    A_1-A_2=\frac{1}{2\pi\lambda}\int_{-\pi}^{\pi}dk_y^\prime\frac{1+\cos{k_y^\prime}}{\sqrt{1-z^2}}\approx-\frac{\sqrt{2}\ln \epsilon}{\pi\lambda}\approx\frac{2}{J_2} \label{A1A2}
\end{equation}
where $\gamma_2=8(\sqrt{2}\ln{(\sqrt{2}+1)}-1)\approx 1.97$. Insert Eq. (\ref{A1A2B}) and Eq. (\ref{A1A2}) to Eq. (\ref{Hessian}) one finds, to the lowest order, the determinant of Hessian vanishes when
\begin{equation}
    \sqrt{\epsilon} \ln{\epsilon}=\frac{\sqrt{\gamma}}{(1-\frac{2}{\eta})}.
\end{equation}
At large $S$ and small $\eta$, the critical spin $S_{\text{$\mathbb{Z}_2$}}$ is hence given by Eq. (\ref{Sch}) in the main text.

For the $J_{2x}=J_{2y}$ case (denoted as the spin nanostrip), one could find that its phase diagram is similar to the spin nanotube case, except for the absence of valence bond solid (VBS) phase. The vanishing VBS phase can be understood by noting that, for general $J_{2x}\geq J_{2y}$, $\Delta_y$ becomes nonzero when $S\geq S_{cl}=(\sqrt{J_{2x}/J_{2y}}-1)/2$, and when $J_{2x}=J_{2y}$ $S_{cl}$ is exactly zero. Through a similar calculation used to compute the $J_{2x}=2J_{2y}$ case, one obtains Eq. (\ref{Sch2}) in the main text, and the derivation is not shown here.

\subsubsection{$N_x\geq4$ case}
For general $N_x$ one can numerically confirm that when $N_x \ \text{mod}\ 4=0$ the lowest energy bosonic state is the one with periodic boundary condition (or translationally invariant ansatz), while if $N_x \ \text{mod}\ 4=2$, the lowest energy state corresponds to the one with anti-periodic boundary condition (or ansatz that lacks translational symmetry). Therefore $k_x$ takes value $\frac{2\pi}{N_x}m+\frac{\pi}{2}$, consistent with the special case $N_x=2$. This periodic boundary condition indicates that system can lower its energy by having those quasi-gapless subbands at transverse momentum $k_x=\pi/2,3\pi/2$. Similar to the $N_x=2$ case, one could first find the critical spin $S_\text{cl}$ where $\Delta_y$ develops and then look for the critical spin $S_{\text{$\mathbb{Z}_2$}}$ where $\mathbb{Z}_2$ phase transition occurs. However, for $N_x\geq 4$, it turns out that $\Delta_x$ and $\Delta_y$ are both present for any finite $S$ if $\eta$ is small, as a consequence of that $J_2$ along $x$-direction is no longer doubled. To see this, one can set $S=0^+$, solve the mean field equations (\ref{MFS})-(\ref{MFDeltay}) and obtain
\begin{equation}
    \lambda_0=\lim_{S\rightarrow 0}\lambda=\frac{J_2}{2N_x}\sum_{k_x} \sin^2k_x=\left\{\begin{array}{cc}
        J_2/2 ,& N_x=2 \\
         J_2/4,& N_x\geq 4\\
    \end{array}\right..
\end{equation}
The Hessian at $S=0^+$ is diagonal and given by
\begin{equation}
        \frac{1}{N_s}\frac{\partial^2 E_g}{\partial \alpha_i\alpha_j}=\begin{pmatrix}
      0 & & & &\\
       & 0 & & &  \\
       & &\frac{8}{J_2}-\frac{2}{\lambda_0} & &\\
        & & & -\frac{8}{J_1}&\\
        & & & &\frac{8}{J_1}-\frac{2}{\lambda_0}\\
    \end{pmatrix}_{ij}.
\end{equation}
Thus for $N_x\geq 4$, $\frac{\partial^2 E_g}{\partial \alpha_2^2}=\frac{\partial^2 E_g}{\partial \alpha_3^2}=0$ when $S=0$, which indicates that $\Delta_x$ and $\Delta_y$ acquire expectation value simultaneously at infinitesimal $S$ when $\eta<1/2$. When $\eta>1/2$,  note that $\frac{\partial^2 E_g}{\partial \alpha_5^2}$ becomes negative, indicating that $\Delta_+$ will first develop while other mean-field parameters except $\lambda$ remain zero at small $S$. We denote this phase as the decoupled antiferromagnetic chains (DAC) and show it in Fig. \ref{fig:Nx4diagram}.

For our purpose we focus on the $\eta<1/2$ case and solve for the critical spin $S_{\text{$\mathbb{Z}_2$}}$ where $\mathbb{Z}_2$ order phase transition occurs. When $\eta$ is close to $1/2$, the $\mathbb{Z}_2$ phase transition is expected to occur at small $S$ where the bosonic gap is large compared to $1/N_x$. In the other words, the phase transition occurs when the Heisenberg correlation length $\xi$ is much smaller than the width $N_x$ of the nanotube, which corresponds to the two dimensional limit. In this case, all the sums over $k_x$ can be converted to integrals as functions such as $1/\omega$ are smooth, which is equivalent to setting $N_x=\infty$. The critical $S_{\text{$\mathbb{Z}_2$}}$ can be solved numerically by letting the determinant of Hessian Eq. (\ref{Hessian45}) be zero and the result is shown as the red dotted line in Fig. (\ref{fig:Nx4diagram}).

At small $\eta\approx 0$, the $\mathbb{Z}_2$ phase transition occurs at large $S$ as in the $N_x=2$ case, corresponding to the one dimensional limit as $\xi\gg N_x$. The mean field equations are similar to the $N_x=2$ case, except that there are also contributions from gapped subbands:
\begin{equation}
    2S+1=\frac{1}{N_x}\left[2\int_{-\pi}^{\pi}\frac{dk_y^\prime}{2\pi}\frac{1}{\sqrt{1-z^2}}+I_S\right], \label{MFSNx4}
\end{equation}
\begin{equation}
    \frac{4}{J_2}\Delta_x=\frac{1}{N_x}\left[2\int_{-\pi}^{\pi}\frac{dk_y^\prime}{2\pi}\frac{u}{\sqrt{1-z^2}}+I_x\right], \label{MFDeltaxNx4}
\end{equation}
\begin{equation}
    \frac{4}{J_2}\Delta_y=\frac{1}{N_x}\left[2\int_{-\pi}^{\pi}\frac{dk_y^\prime}{2\pi}\frac{u\cos k_y^\prime}{\sqrt{1-z^2}}+I_y\right], \label{MFDeltayNx4}
\end{equation}
where at large $S$
\begin{equation}
    I_s\approx \sum_{k_x^\prime\neq0,\pi}\int_{-\pi}^{\pi}\frac{dk_y^\prime}{2\pi}\frac{1}{\sqrt{1-\tilde{z}^2}},
\end{equation}
\begin{equation}
    I_x\approx\sum_{k_x^\prime\neq0,\pi}\int_{-\pi}^{\pi}\frac{dk_y^\prime}{2\pi}\frac{\tilde{u}\cos{k_x}}{\sqrt{1-\tilde{z}^2}},
\end{equation}
\begin{equation}
    I_y\approx\sum_{k_x^\prime\neq0,\pi}\int_{-\pi}^{\pi}\frac{dk_y^\prime}{2\pi}\frac{\tilde{u}\cos{k_y^\prime}}{\sqrt{1-\tilde{z}^2}},
\end{equation}
where we define $k_x^\prime=k_x-\pi/2$ and $\tilde{z}=(\cos{k_x^\prime}+\cos{k_y^\prime})/2$. Combine the above equations one obtains
\begin{align}
    2S+1 &\approx\frac{1}{N_x\pi}[\sqrt{2}(5\ln{2}-\ln{\epsilon})+\pi I_S],\\\nonumber
    \quad \lambda&\approx -\frac{\sqrt{2}J_2}{N_x\pi}\ln{\epsilon}. 
\end{align}
The determinant of the Hessian can be determined in a manner similar to the $N_x=2$ case:
\begin{equation}
    A_1+A_2-2B\approx \frac{8}{J_2\epsilon \ln \epsilon},
\end{equation}
\begin{equation}
    A_1-A_2\approx\frac{2}{J_2},
\end{equation}
and
\begin{align}
    &A_1+A_2+2B\\\nonumber
    &=-\frac{1}{N_x}\left[\frac{2}{\lambda}\int_{-\pi}^{\pi}\frac{dk_y^\prime}{2\pi}\frac{(1-u)(1+\cos{k_y^\prime})}{(u+1)\sqrt{1-u^2}}+\frac{2I_-}{\lambda}\right]\\\nonumber
    &=-2\frac{\gamma/2\sqrt{2}\pi+I_-}{N_x\lambda},
\end{align}
where 
\begin{equation}
    I_-\approx\sum_{k_x^\prime\neq0,\pi}\int_{-\pi}^{\pi}\frac{dk_y}{2\pi}\frac{(1-\tilde{u})(1+\cos{k_y^\prime})}{2(\tilde{u}+1)\sqrt{1-\tilde{u}^2}}.
\end{equation}
Combine all the above equations with Eq. (\ref{Hessian}) one obtains
\begin{multline}
    S_{\text{$\mathbb{Z}_2$}}\exp\left(-\frac{N_x\pi(S_{\text{$\mathbb{Z}_2$}}+1/2)}{\sqrt{2}}\right)\\=\exp\left(-\frac{\pi}{2\sqrt{2}}I_s\right)\frac{\sqrt{\gamma+2\sqrt{2}\pi I_-}}{16N_x\pi }\eta. \label{Nx4Sch}
\end{multline}

\section{Scaling of the spin gap and string order parameters}\label{appendix:scaling}
\begin{figure}[tbh]
    \centering
    \includegraphics[width=\linewidth]{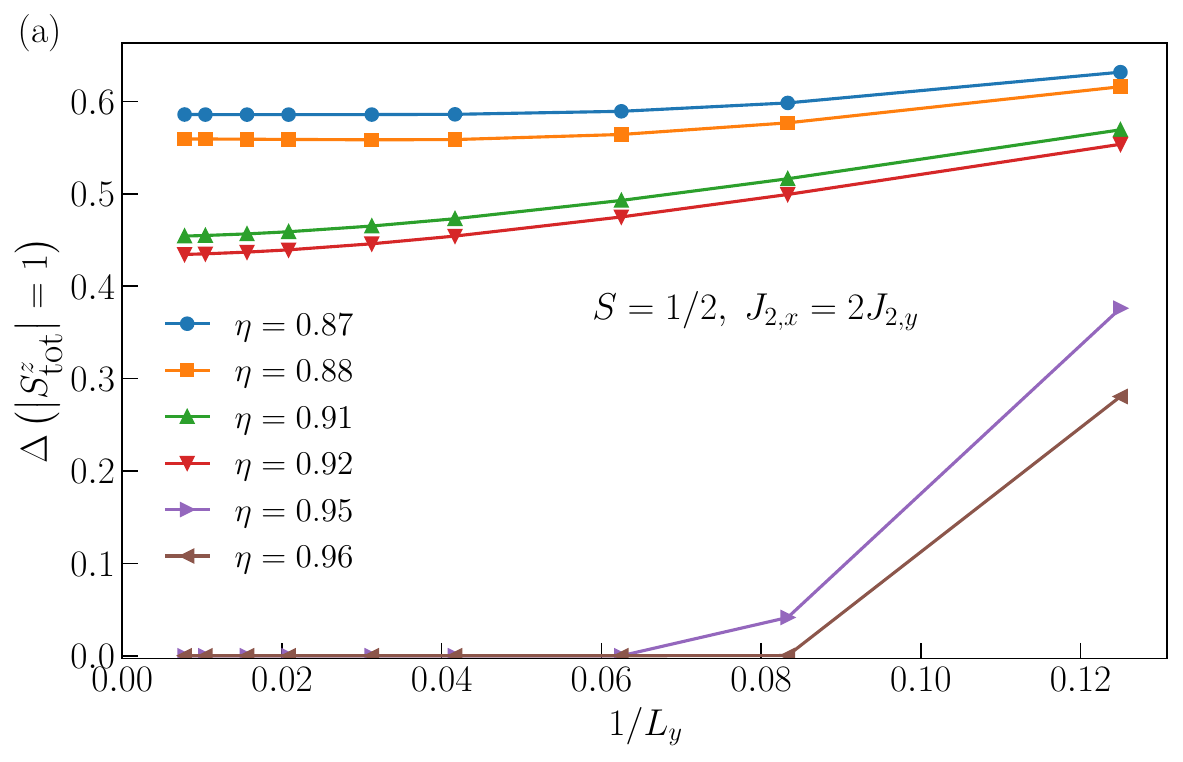}
    \includegraphics[width=\linewidth]{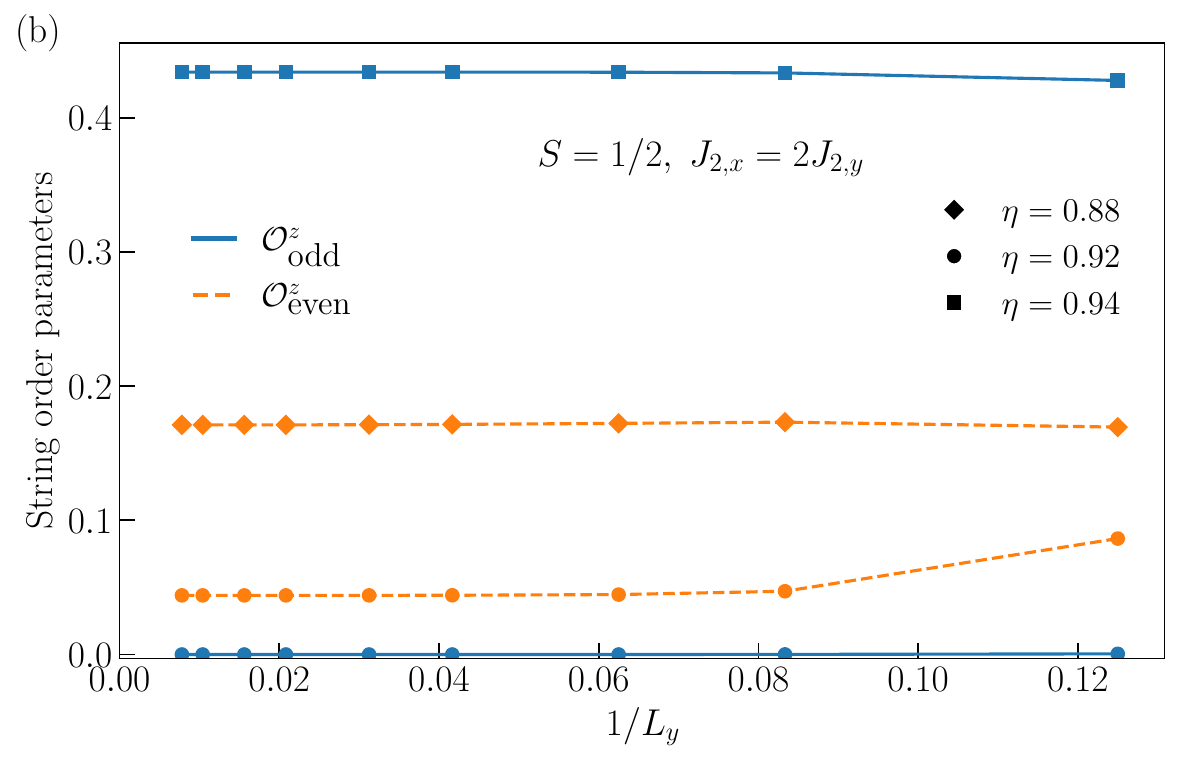}
    \caption{(a) Scaling of the spin gap $\Delta\left(\left|S^z_\textrm{tot}=1\right|\right)$ and (b) the string order parameters $\mathcal{O}^z_\textrm{odd}$ and $\mathcal{O}^z_\textrm{even}$ (lower panel) with the nanotube length $L_y$ for several values of $\eta=J_1/2J_2$. The data corresponds to the case where $J_{2x} = 2 J_{2y}$.}
    \label{fig:dmrg-scaling}
\end{figure}

For the sake of completeness, we show here a few examples of the scaling of the spin gap and string order parameters with respect to the system size. In Fig.~\ref{fig:dmrg-scaling}\hyperlink{fig:dmrg-scaling}{(a)} we have selected several $\eta=J_1/2J_2$ values, in order to cover all three phases, to present the scaling of the spin gap $\Delta\left(\left|S^z_\textrm{tot}\right|=1\right)$ as a function of the inverse of the nanotube length $L_y$. Here we have focused only on the nanotube setup, as we are able to reach larger system sizes than the nanostrip case. Moreover, we should point out that the scaling plot for the energy gap between the first-excited and ground states of sector $S^z_\textrm{tot}=0$ does not behave as smoothly as the spin gap represented in Fig.~\ref{fig:dmrg-scaling}\hyperlink{fig:dmrg-scaling}{(a)}. The reason lies in a poor choice of the weight factor $w$ of the penalized Hamiltonian, which we have set in our preliminary calculations as $w=10$ regardless of system size and $\eta$ value. However, by tuning the parameter $w$, and also by adding noise to the initial sweeps of the DMRG method, we have verified that one can obtain more consistent results for the energy gap.

Similarly, Fig.~\ref{fig:dmrg-scaling}\hyperlink{fig:dmrg-scaling}{(b)} displays the scaling of the string order parameters $\mathcal{O}^z_\textrm{odd}$ and $\mathcal{O}^z_\textrm{even}$ (Eqs.~\ref{eq:odd_string_op} and~\ref{eq:even_string_op}, respectively). For each string order parameter, we have chosen two $\eta$ values, one within the $\mathbb{Z}_2$-helical phase and the other within the phase where the corresponding parameter is finite. Notice that the string order parameters saturate for relatively small system sizes, thus indicating that the data for the largest system, namely $L_y=128$, is a good approximation to the thermodynamic limit.

%

\end{document}